\documentclass
[preprint, showpacs,preprintnumbers,amsmath,amssymb,nofootinbib]{revtex4}
\usepackage [latin1]{inputenc}
\usepackage{subfig}
\usepackage{amssymb}
\usepackage{natbib}
\usepackage{graphicx}
\usepackage{dcolumn}
\usepackage{bm}
\renewcommand\theequation{\arabic{equation}}
\newcommand{\bea}{\begin{eqnarray}}
\newcommand{\eea}{\end{eqnarray}}

\begin{document}
\title{\Large \bf Impact of Interacting Dark Energy on the Growth of Matter Density Perturbations: Observational Constraints from DESI and Multi-Probe Data}
\author{Fan Yang$^1$, Rongrong Zhai$^2$, Xiangyun Fu$^1$\footnote{corresponding author: xyfu@hnust.edu.cn}, Bing Xu$^3$\footnote{corresponding author: xub@ahstu.edu.cn},
Kaiwen Liu, Chikun Ding$^4$, and Yang Huang$^1$}
\address{$^1$Department of Physics, Key Laboratory of Intelligent Sensors and Advanced Sensor Materials, Hunan University of Science and Technology, Xiangtan, Hunan 411201, China\\
$^2$Department of Physics, Xinzhou Normal University, Xinzhou 034000, Shanxi, China\\
$^3$School of Electrical and Electronic Engineering, Anhui Science and Technology University, Bengbu, Anhui 233030, China\\
$^4$ Department of Physics, Huaihua University, Huaihua, 418000, China}

\begin{abstract}
	We investigate the impact of a non-gravitational dark sector interaction on the growth of matter density perturbations within both the interacting $w$CDM and the dynamical Chevallier-Polarski-Linder (CPL) scenarios. For $w$CDM model, we develop a parameterization for the growth rate based on a second-order approximation for the growth index $\gamma$ that explicitly includes the coupling constant $\alpha$. Our analysis reveals a theoretical degeneracy: the coupling induces a correction $\Delta\gamma \simeq 1.1\alpha$ in both models, allowing an interacting dark energy model to mimic the growth index predicted by certain modified gravity theories. Then, we confront the models with the latest multi-probe observations, including the Pantheon+ sample of Type Ia supernovae, Baryon Acoustic Oscillation (BAO) data from the Sloan Digital Sky Survey (SDSS) and the second data release (DR2) of the Dark Energy Spectroscopic Instrument (DESI), Cosmic Microwave Background (CMB) measurements, Hubble parameter $H(z)$ data, and redshift-space distortion (RSD) measurements. Our analysis finds that the coupling constant is consistent with zero at approximately the $3\sigma$ and $2\sigma$ confidence levels for $w$CDM and CPL models, respectively, showing no definitive statistical evidence for a departure from the standard $\Lambda$CDM cosmology. The observational constraints strongly disfavor the region of parameter space where interacting dark energy can mimic modified gravity, restricting the growth index to a common approximate interval of $0.53 \lesssim \gamma \lesssim 0.60$ for both models. This reinforces the growth index as a robust diagnostic for distinguishing between a non-minimal interaction in the dark sector and a genuine modification of gravity with current data.
\end{abstract}


 \maketitle
 \renewcommand{\baselinestretch}{1.5}
\section{Introduction}\label{sec1}
The current accelerating expansion of the Universe presents a profound challenge to the standard model of cosmology and fundamental physics~\cite{Peebles2003,Padmanabhan,AGRiess,Riess,pastier,dnspergel,mtegmark}. Within the framework of General Relativity (GR), it is attributed to a dominant component with negative pressure, dubbed dark energy (DE). The $\Lambda$ cold dark matter ($\Lambda$CDM) model, in which DE is described by a cosmological constant $\Lambda$, provides an excellent fit to a wide range of cosmological data. However, it faces persistent theoretical challenges, such as the coincidence problem, and observational tensions---most notably the Hubble tension, i.e., the discrepancy in the Hubble constant $H_0$ between early and late universe probes~\cite{Valentino2021}. These issues motivate the exploration of dynamical alternatives.

A well-studied class of models introduces a non-gravitational interaction between dark matter (DM) and DE, known as interacting dark energy (IDE)~\cite{Amendola2000,JHHe2008,He2009,Martinelli2010,Chimento2010,Pourtsidou2013,Forconi2024,Halder2024,Benisty2024,Zhu2026}. Such interactions can alter both the expansion history and the growth of cosmic structure, offering potential avenues to alleviate cosmological tensions~\cite{Yang2021,Sandro2009,Amendola2000}. While earlier works confronting such models with various observational data often found no statistically significant deviation from the standard non-interacting $\Lambda$CDM paradigm~\cite{Dalal2001,Guo2007,Chen2010,Cao2011,Wang2022,Nong2024,Wang2024,Cao2025,Halder2024,Giare2024,Li2026,Zheng2022,Hou2023,Li2024}, recent high-precision measurements have begun to shift this landscape. In parallel, modifications to GR on cosmological scales (for a review see Ref.~\cite{Caldwell2009})---such as DGP braneworld gravity~\cite{Dvali2000} or $f(R)$ theories---provide another compelling explanation for cosmic acceleration, often predicting background expansion histories identical to those of DE models. Distinguishing between these fundamentally different mechanisms---a dark-sector interaction versus a modified gravity theory---requires probes beyond the Hubble flow.

The growth of linear matter density perturbations serves as a powerful discriminant among cosmological models, as it directly probes the underlying gravitational physics and the dynamics of the cosmic components driving the expansion history. A key observable is the growth rate $f \equiv d\ln\delta / d\ln a$, where $\delta \equiv \delta\rho_{\rm m} / \rho_{\rm m}$ is the linear matter density contrast, $a$ is the scale factor related to the cosmic redshift $z$ by $a = (1+z)^{-1}$. To facilitate comparisons with observations and to encapsulate the theoretical dependence of $f$ on cosmology, a widely adopted and phenomenologically useful parameterization is given by $f(z) \simeq \Omega_{\rm m}(z)^{\gamma(z)}$~\cite{jnfry}. Here, $\Omega_{\rm m}(z)$ is the fractional energy density of matter, and $\gamma(z)$ denotes the growth index, which may in general be redshift-dependent. This parameterization is not merely a fitting formula; rather, it emerges naturally from the dynamics of perturbation growth in many scenarios and has been demonstrated to provide an excellent approximation across a broad range of cosmological models, including those with dynamical dark energy and certain modifications of gravity~\cite{Linder2007,Linder2005,Cai2022,Ghedini2024}.

The value of the growth index, particularly its asymptotic or characteristic amplitude, varies across different theoretical frameworks, making it a potent diagnostic tool. Within the standard $\Lambda$CDM model under General Relativity, the growth index is about $\gamma \approx 6/11 \simeq 0.545$. In contrast, modified gravity theories predict different values: for instance, the self-accelerating branch of the DGP braneworld model yields $\gamma \simeq 11/16 \approx 0.6875$, while for most viable $f(R)$ gravity models, $\gamma$ typically falls within the range $0.40$--$0.43$~\cite{dpolarski2008,Tsujikawa2009,Fu2009,Fu2010}. It is worth noting, however, that the growth index in $f(R)$ gravity is generally scale-dependent and varies among specific models. For example, in the Starobinsky model, $\gamma(z) = 0.399 - 0.246z$ for $\Omega_{\mathrm{m},0}=0.315$ \cite{Gannouji2009,starobinsky2007}, whereas in the Hu-Sawicki model, $\gamma(a) = 0.753 + 0.690(1-a)$ on a different scale \cite{Basilakos2017}. Here, the  $\Omega_{\mathrm{m},0}$ denotes the present--day fractional energy density of matter. This intermediate value among $\Lambda$CDM, DGP, and  $f(R)$ demonstrates that even models with identical background evolution can be distinguished through their perturbation growth, highlighting the necessity of probing structure formation to break degeneracies among competing theories. Consequently, a precise measurement of $\gamma$ could, in principle, distinguish between a cosmological constant supported by General Relativity and modifications of the gravitational law on large scales. In addition, the redshift evolution of $\gamma(z)$ provides additional leverage, as different physical mechanisms (modified gravity vs. interaction) may imprint distinct trajectories on $\gamma(z)$ even if they predict similar values at $z=0$~\cite{dpolarski2008,Fu2009,Fu2010}. Consequently, the growth index serves as a potent tool for distinguishing dark energy from modified gravity.

A non-gravitational coupling between dark matter and dark energy also modifies the growth of cosmic structure~\cite{Amendola2000,JHHe2008,He2009,Martinelli2010,lpchimento1997,jdbarrow2006}. Indeed, it has been demonstrated that an IDE model can be constructed to reproduce both the expansion history and the growth history of a given modified gravity scenario, rendering the two classes of models indistinguishable based solely on these two observational probes. A notable theoretical example of this degeneracy was provided by \cite{Wei2008}, who demonstrated that within the framework of interacting quintessence---where dark energy is described by a canonical scalar field $\phi$ with an interaction of the form $Q \propto Q(\phi)\rho_m \dot{\phi}$---one can reconstruct a model that exactly reproduces both the expansion history $H(z)$ and the linear growth history $\delta(z)$ of the DGP model. This implies that, in principle, the combination of expansion and growth probes alone may not be sufficient to distinguish between IDE and modified gravity. This degeneracy highlights the necessity for additional observational probes to discriminate between IDE and modified gravity. Furthermore, recent analyses of combined cosmological datasets, including the Dark Energy Spectroscopic Instrument (DESI) DR2 \cite{Abdul-Karim2025} and the re-calibrated DES supernova sample~\cite{Popovic2025}, have reported a statistical preference for a time-varying dark energy equation of state over the cosmological constant $\Lambda$. Consequently, IDE models have regained considerable interest, because the energy transfer between dark matter and dark energy naturally provides a mechanism for an effective dynamical dark energy equation of state \cite{Guedezounme2025,Westhuizen2025a,Westhuizen2025b}.

Motivated by these considerations, we investigate the interacting dark energy scenario considering both the $w$CDM model with a constant equation of state and the dynamical Chevallier-Polarski-Linder (CPL) parameterization, $w(a) = w_0 + w_a(1-a)$. We focus on a coupling proportional to the dark energy density, $Q=\alpha H\rho_{\rm d}$ (where $H$ is the Hubble parameter and $\rho_{\rm d}$ the dark energy density). This specific form is widely adopted because it is mathematically tractable, avoids unphysical negative dark energy densities at high redshifts, and naturally emerges in scalar-tensor or holographic frameworks \cite{Wang2016,Westhuizen2024}. To ensure the rigorous accuracy of background evolution at high redshifts---crucial for CMB distance priors---we explicitly incorporate the radiation component into our equations. This context motivates three key tasks: (i) a quantitative assessment of how the coupling parameter $\alpha$ influences the growth rate $f$; (ii) an evaluation of the accuracy of the standard parameterization $f \simeq \Omega_{\rm m}^{\gamma}$ in the coupled case, and the development of an improved, computationally efficient parameterization that explicitly incorporates $\alpha$; and (iii) confronting both interacting $w$CDM and CPL models with the latest multi-probe cosmological data to test whether the theoretical degeneracy identified by \cite{Wei2008} persists under precise observational scrutiny. Our aim is to quantify the extent to which the interaction-induced correction shifts the growth index within these interacting frameworks, and to establish the growth index as a robust diagnostic for distinguishing between a dark-sector interaction and a genuine modification of gravity. These considerations form the primary motivation for the present work.

The structure of this paper is as follows. In Section II, we present the background evolution equations including the radiation component for both the interacting $w$CDM and CPL models. Section III is devoted to the growth of matter perturbations, where we derive the modified growth equation and obtain a new second-order approximation for the growth index $\gamma(z)$. In Section IV, we describe the diverse cosmological datasets employed in our analysis and the statistical methodology. Our main results from the Markov Chain Monte Carlo (MCMC) constraints, including a comprehensive comparative analysis of both $w$CDM and CPL scenarios, are presented and discussed in Section V. Finally, we summarize our conclusions in Section VI.

\section{Background evolution}\label{sec2}
In a spatially flat Friedmann-Lema$\hat{\rm{\i}}$tre-Robertson-Walker universe, we consider interacting dark energy models where cold dark matter and dark energy interact with each other. To accurately compute the background evolution at high redshifts (which is crucial for CMB data), we explicitly include the radiation component. The background equations of motion are given by
\begin{equation}
	\label{friedmanne} {H^2}=\frac{8\pi G}{3}(\rho_{\rm m}+\rho_{\rm d}+\rho_{\rm r}),
\end{equation}
and
\begin{equation}
	\label{friedmanne2} {\dot H\over H^2}=-\frac{3}{2}\left[\Omega_{\rm m}+(1+w(a))\Omega_{\rm d}+\frac{4}{3}\Omega_{\rm r}\right]\,.
\end{equation}
Here, $G$ is Newton's gravitational constant; $\rho_{\rm r}$ is the energy density of radiation, while $\Omega_{\rm d}$ and $\Omega_{\rm r}$ denote the fractional energy densities of dark energy and radiation, respectively. The continuity equations for the three components are
\begin{eqnarray}
	\label{contimuitye}
	&&\dot{\rho}_{\rm m}+3H\rho_{\rm m}=\alpha H\rho_{\rm d}\,,\\
	&&\dot{\rho}_{\rm d}+3H{\rho}_{\rm d}(1+w(a))=-\alpha H\rho_{\rm d}\,,\\
	&&\dot{\rho}_{\rm r}+4H\rho_{\rm r}=0\,.
\end{eqnarray}
Here, $\dot{\rho}$ denotes differentiation with respect to cosmic time. In this work, we investigate two specific scenarios for the dark energy equation of state: the $w$CDM model with a constant equation of state ($w(a) = w$), and the dynamical Chevallier-Polarski-Linder (CPL) parameterization, $w(a) = w_0 + w_a(1-a)$.

Solving the above equations (the detailed step-by-step derivations for both models are provided in Appendix~\ref{app:derivation}), we obtain the normalized Hubble parameter $E(z)^2 \equiv H^2/H_0^2$ and the fractional energy density of matter $\Omega_{\rm m}(z)$.

For the interacting $w$CDM model, they are given by:
\begin{eqnarray}
	\label{eze_wcdm}
	E(z)^2&=&\bigg(\Omega_{\rm m,0}+{\alpha\Omega_{\rm d,0}\over 3w+\alpha}\bigg)(1+z)^3-{\alpha\Omega_{\rm d,0}\over 3w+\alpha}(1+z)^{3(1+w)+\alpha}\nonumber\\
	&&+\Omega_{\rm d,0}(1+z)^{3(1+w)+\alpha} + \Omega_{\rm r,0}(1+z)^4\,,\\
	\label{omegam_wcdm}
	\Omega_{\rm m}(z)&=&\frac{\big(\Omega_{\rm m,0}+{\alpha\Omega_{\rm d,0}\over 3w+\alpha}\big)(1+z)^3-{\alpha\Omega_{\rm d,0}\over 3w+\alpha}(1+z)^{3(1+w)+\alpha}}{E(z)^2}\,.
\end{eqnarray}

For the interacting CPL model, the analytical integration yields:
\begin{eqnarray}
	\label{eze_cpl}
	E(z)^2 &=& \Omega_{\rm r,0}(1+z)^4 + \Omega_{\rm m,0}(1+z)^3 + \Omega_{\rm d,0}(1+z)^{3(1+w_0+w_a)+\alpha} e^{3w_a(\frac{1}{1+z}-1)} \nonumber \\
	&& + \alpha \Omega_{\rm d,0} (1+z)^3 e^{-3w_a} \int_1^{\frac{1}{1+z}} x^{-(1+3w_0+3w_a+\alpha)} e^{3w_a x} dx \,,\\
	\label{omegam_cpl}
	\Omega_{\rm m}(z) &=& \frac{\Omega_{\rm m,0}(1+z)^3 + \alpha \Omega_{\rm d,0} (1+z)^3 e^{-3w_a} \int_1^{\frac{1}{1+z}} x^{-(1+3w_0+3w_a+\alpha)} e^{3w_a x} dx}{E(z)^2} \,.
\end{eqnarray}

In both cases, the present-day fractional energy density of dark energy is $\Omega_{\rm d,0} = 1 - \Omega_{\rm m,0} - \Omega_{\rm r,0}$. The present-day fractional energy density of radiation $\Omega_{\rm r,0}$ is determined by the relation $\Omega_{\rm r,0} = \Omega_{\rm m,0} / (1 + z_{\rm eq})$, where the redshift of matter-radiation equality is given by $z_{\rm eq} = 2.5 \times 10^4 \Omega_{\rm m,0} h^2 (T_{\rm cmb} / 2.7\,{\rm K})^{-4}$, with the CMB temperature $T_{\rm cmb} = 2.7255\,{\rm K}$~\cite{Shen2025}.

The general differential equation governing the evolution of the fractional energy density of matter is:
\begin{equation}
	\label{eq:dOmega_m_full}
	{d\Omega_{\rm m}\over d\ln a}=\alpha\Omega_{\rm d} + 3w(a)\Omega_{\rm m}\Omega_{\rm d} + \Omega_{\rm m}\Omega_{\rm r}\,.
\end{equation}
These full expressions are strictly used for calculating the background observational quantities (SNIa, $H(z)$, BAO, and CMB).

However, for the analysis of the growth of matter perturbations ($f\sigma_8$ data) which occurs in the late universe ($z \lesssim 2$), the radiation density is entirely negligible ($\Omega_{\rm r} \to 0$, and thus $\Omega_{\rm d} \to 1 - \Omega_{\rm m}$). In this late-time limit, Eq.~\eqref{eq:dOmega_m_full} naturally reduces to the following simplified form:
\begin{eqnarray}
	&&{d\Omega_{\rm m}\over d\ln a} \simeq 3w(a)\Omega_{\rm m}(1-\Omega_{\rm m})+\alpha(1-\Omega_{\rm m})\,.\label{eq:dOmega_m}
\end{eqnarray}
We will use this reduced form in the next section to derive the modified differential equations for the growth rate.


\section{Growth of Matter Perturbations in IDE Model}

In the sub-horizon limit, due to the large sound speed of dark energy, the dark energy perturbations are significantly suppressed in comparison to the dark matter perturbations~\cite{jackson0901,He2009,caldera-cabral}. Therefore, the effect of dark energy perturbations on the growth of DM density perturbations can be neglected. To linear order of perturbation, at large scales the matter density perturbation $\delta$ satisfies the following equation~\cite{wangli1998,Fu2009,Fu2010},
\begin{equation}
	\label{denpert} \ddot{\delta}+2H\dot{\delta}-4\pi
	G\,\rho_{\rm m}\delta=0,
\end{equation}
where the dot denotes the derivative with respect to time $t$. Defining the growth rate as $f \equiv d\ln\delta/d\ln a$, and substituting it into Eq.~\eqref{denpert}, we obtain
\begin{equation}
	\label{grwthfeq1} {d\; f\over d\ln
		a}+f^2+\left(\frac{\dot{H}}{H^2}+2\right)f=\frac{3}{2}\Omega_{\rm m}.
\end{equation}
In general, there is no analytical solution to Eq.~\eqref{grwthfeq1}, and one needs to solve it numerically.

For the case of the linear order of matter perturbation with the coupling between dark matter and dark energy, the growth function $\delta$ at a much smaller scale than the Hubble radius satisfies the following equation~\cite{jackson0901,He2009,caldera-cabral},
\begin{eqnarray}
	\label{gamma-2}
	\ddot{\delta}+2H(1+\alpha{1-\Omega_{\rm m}\over \Omega_{\rm m}})\dot{\delta} +\alpha{1-\Omega_{\rm m}\over
		\Omega_{\rm m}}[H^2+\dot{H}-H^2(\alpha+3w(a)-1)]\delta=4\pi G\rho_{\rm m} \delta.
\end{eqnarray}
Substituting Eq.~\eqref{eq:dOmega_m} into Eq.~\eqref{gamma-2}, we get the modified differential equation for the growth rate $f$:
\begin{eqnarray}
	\label{wcdmfeq2}
	&&[3w(a)\Omega_{\rm m}(1-\Omega_{\rm m})+\alpha(1-\Omega_{\rm m})]\frac{df}{d\Omega_{\rm m}}+f^2+\bigg[{\dot{H}\over H^2}+2\bigg(1+{\alpha(1-\Omega_{\rm m})\over{\Omega_{\rm m}}}\bigg)\bigg]f\nonumber\\&&+{\alpha(1-\Omega_{\rm m})\over{\Omega_{\rm m}}}
	\bigg(2+{\dot{H}\over H^2}-\alpha-3w(a)\bigg)=\frac{3}{2}\Omega_{\rm m}\,.
\end{eqnarray}

\begin{figure}[htbp]
	\centering
	\includegraphics[width=0.48\textwidth]{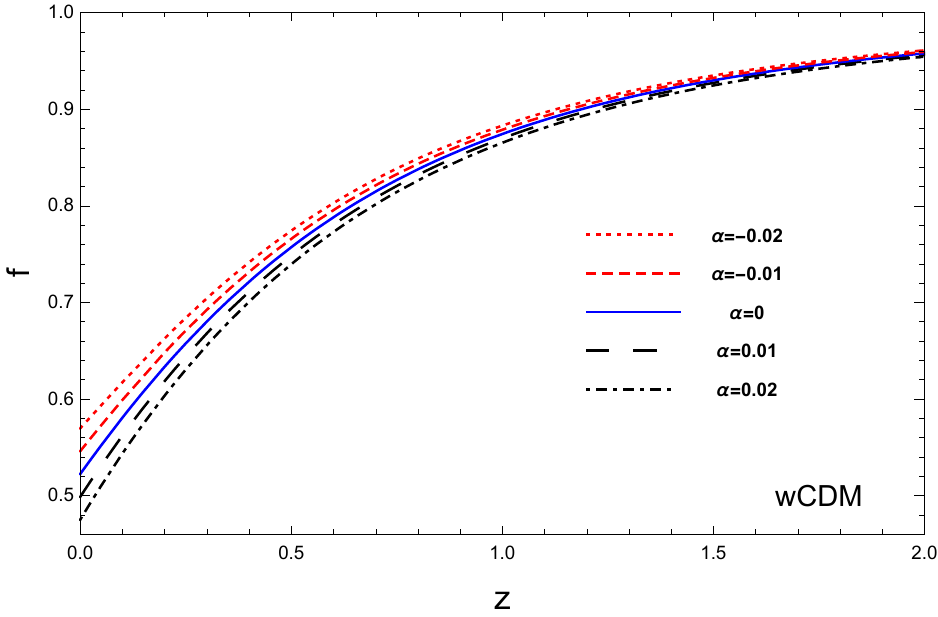}
	\includegraphics[width=0.48\textwidth]{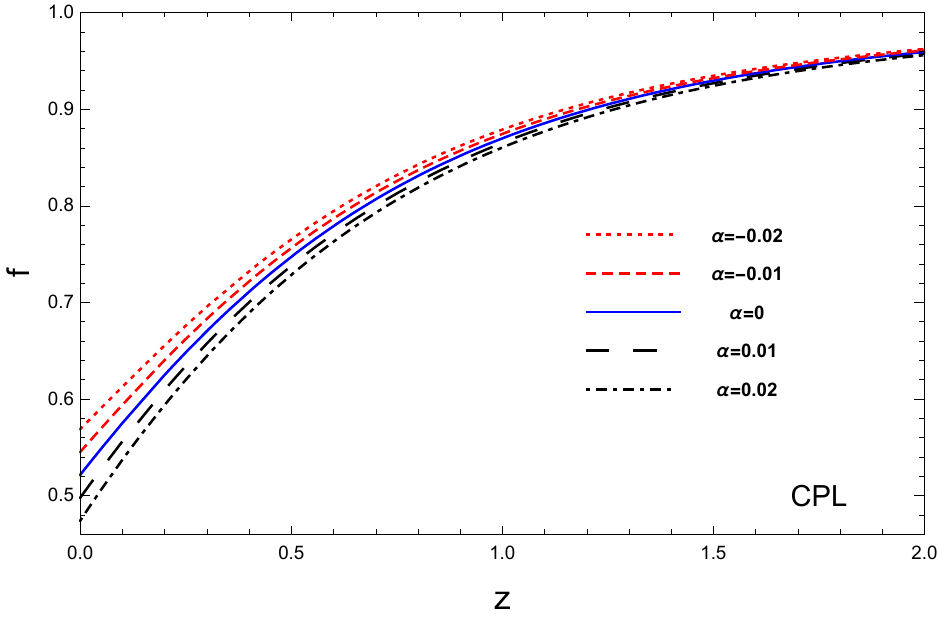}
	\caption{The evolution of the growth rate $f(z)$ with respect to redshift $z$ for different values of the coupling constant $\alpha$. Left panel (a): The interacting $w$CDM model with $\Omega_{\rm m,0}=0.31$ and $w=-1$. Right panel (b): The interacting CPL model with $\Omega_{\rm m,0}=0.31$, $w_0=-0.9$, and $w_a=-0.28$.}
	\label{fig:fz_evolution}
\end{figure}

This equation forms the basis of our first analysis method, which is applicable to both the interacting $w$CDM and CPL models. The numerical results from Eq.~\eqref{wcdmfeq2} are shown in Fig.~\ref{fig:fz_evolution}, demonstrating that a positive coupling constant corresponds to a smaller growth rate $f$ compared to the scenario with no coupling, and vice versa. The left panel (Fig.~\ref{fig:fz_evolution}a) illustrates the interacting $w$CDM model, while the right panel (Fig.~\ref{fig:fz_evolution}b) shows the interacting CPL model.

\begin{figure}[htbp]
	\centering
	\includegraphics[width=0.48\textwidth]{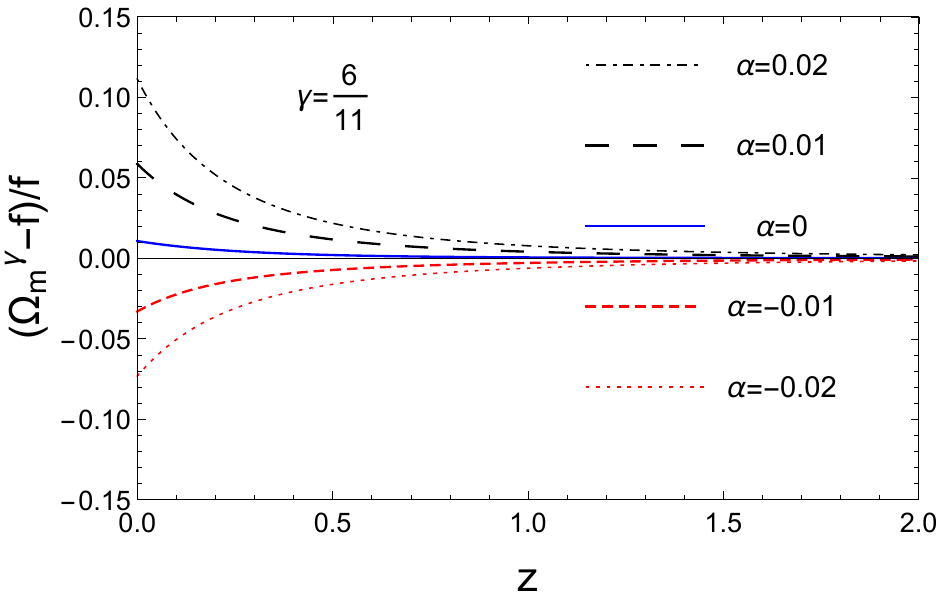}
	\includegraphics[width=0.48\textwidth]{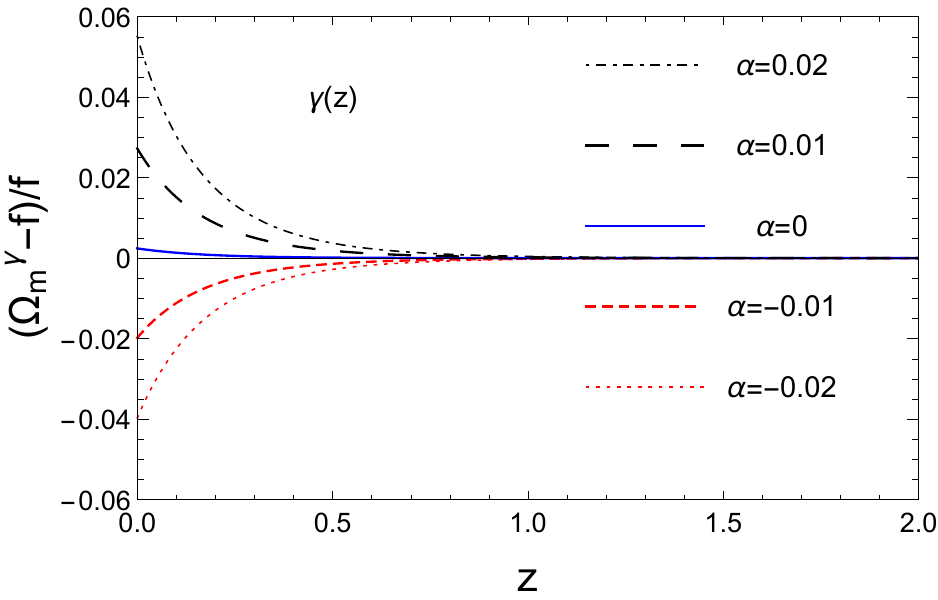}
	\caption{The relative difference between the numerical growth rate $f$ and the parameterization $\Omega_{\rm m}^\gamma$ for the interacting $w$CDM model ($\Omega_{\rm m,0}=0.31, w=-1$). Left panel (a): Using a constant growth index $\gamma = 6/11$. Right panel (b): Using the improved second-order approximation for $\gamma(z)$.}
	\label{fig:error_wcdm}
\end{figure}

Our second analysis method utilizes the parameterized form $f \approx \Omega_{\rm m}^\gamma$, which is initially derived for the interacting $w$CDM model. We first consider the $w$CDM dark energy model without interaction. In this case, Eq.~\eqref{grwthfeq1} can be rewritten as
\begin{equation}
	\label{wcdmfeq}
	[3w\Omega_{\rm m}(1-\Omega_{\rm m})]\frac{df}{d\Omega_{\rm m}}+f^2+\bigg({\dot{H}\over H^2}+2\bigg)f=\frac{3}{2}\Omega_{\rm m},
\end{equation}
making the assumption that the dark energy component can be described by a constant equation of state $w$ and negligible dark energy perturbations. Using the parameterized form $\Omega_{\rm m}^\gamma$ for the growth rate $f$ and expanding Eq.~\eqref{wcdmfeq} around $\Omega_{\rm m}\sim 1$ (good approximation especially at $z\geq 1$), in the case of no interaction term, the growth index $\gamma$ can be approximated as
\begin{equation}
	\label{gamma-3}
	\gamma(z)\simeq{3(1-w)\over 5-6w}+
	\frac{3(1-w)(2-3w)}{2(5-{6w})^2(5-12w)}(1-\Omega_{\rm m}).
\end{equation}
Although this expansion is performed around $\Omega_{\rm m} \sim 1$ (corresponding to the high-redshift matter-dominated epoch), the growth index $\gamma(z)$ is known to be remarkably insensitive to the background evolution. Consequently, the asymptotic solution derived at high redshift remains highly accurate even at low redshifts ($z \lesssim 1$), a robust mathematical property that has been extensively validated in both standard dark energy and modified gravity models \cite{wangli1998,Linder2005,dpolarski2008,Wei2008PLB,Sharma2021}. It is evident that a constant $\gamma$ serves as the zeroth-order approximation to $\gamma(z)$. The relative discrepancy between the parameterization $\Omega_{\rm m}^\gamma$, where the growth index is assumed as a constant, and the theoretical value of the growth rate $f$ is illustrated in Fig.~\ref{fig:error_wcdm}a for the $w$CDM model. We obtain that the relative discrepancy remains below 1\% for the non-interacting case ($\alpha=0$). This discrepancy is much smaller than the average measurement error of the growth rate $f\sigma_8$, which is approximately 15\%~\cite{Howlett2017,Avila2021,Bautista2021,Gil2020} (listed in Table 2 of Ref.~\cite{Huang2022}). Here, $\sigma_8(z)$ is defined as $\sigma_8(z)=\sigma_8(z = 0)\delta(z)/\delta_0 = \sigma_{8}\delta(z)/\delta_0$, where $\delta(z)/\delta_0$ is the redshift-dependent Root Mean Square (RMS) fluctuations of the linear density field within spheres of radius $8h^{-1}{\rm Mpc}$, as described in Ref.~\cite{Nesseris2017}. $h$ is the Hubble constant $H_0$ in units of $100\, {\rm km}\, {\rm s}^{-1}\, {\rm Mpc}^{-1} $. Therefore, the parameterization $\Omega_{\rm m}^\gamma$ with a constant $\gamma$ aligns well with the actual growth rate $f$ in the non-interacting scenarios.

To investigate the impact of this interaction on the growth index, we adopt the parameterization $f \approx \Omega_{\rm m}^{\gamma(z)}$, in which $\gamma(z)$ depends on $\alpha$, $w$ and $z$. Substituting this ansatz into the modified growth equation, Eq.~\eqref{wcdmfeq2}, we obtain the governing equation for the growth index in the IDE scenario:
\begin{eqnarray}
	\label{gamma-4}
	&&[3w\Omega_{\rm m}(1-\Omega_{\rm m})+\alpha(1-\Omega_{\rm m})]\ln \Omega_{\rm m} {d {\gamma(z)}\over d\Omega_{\rm m}}+[3w\Omega_{\rm m}(1-\Omega_{\rm m})+\alpha(1-\Omega_{\rm m})]{{\gamma(z)}\over\Omega_{\rm m}}\nonumber\\
	&&+\Omega^{\gamma(z)}_{\rm m} +\bigg[2\bigg(1+\alpha{1-\Omega_{\rm m}\over \Omega_{\rm m}}\bigg)+{\dot{H}\over H^2}\bigg]+\alpha{1-\Omega_{\rm m}\over{\Omega_{\rm m}^{{\gamma(z)}+1}}}\bigg[1+{\dot{H}
		\over H^2}-(\alpha+3w-1)\bigg]\nonumber\\&&={3\over2}\Omega_{\rm m}^{1-{\gamma(z)}}.
\end{eqnarray}
In the context of interacting dark energy, treating the dark-sector coupling perturbatively is a well-established methodology. Physically, while local gravity tests strictly forbid dark energy from coupling to ordinary matter \cite{caldera-cabral}, the interaction within the dark sector itself must also be sufficiently weak ($|\alpha| \ll 1$) to remain consistent with stringent cosmological constraints from the cosmic microwave background and large-scale structure observations \cite{Giare2024}. Methodologically, employing a small-coupling approximation or a Taylor expansion for the growth index is a standard approach to maintain analytical tractability while accurately capturing the leading-order phenomenological signatures of the interaction \cite{Sharma2021}. Previous studies have demonstrated that under the weak coupling assumption, such analytical approximations are in excellent agreement with exact numerical solutions at low redshifts. Therefore, expanding Eq.~\eqref{gamma-4} around $\Omega_{\rm m} \sim 1$ and $\alpha \sim 0$ to second order yields our improved growth index for the $w$CDM model:
\begin{eqnarray}
	\label{gamma-5}
	\gamma(z)\simeq{3(1-w)\over 5-6w}+a_1(1-\Omega_{\rm m})+a_2\alpha+b_1(1-\Omega_{\rm m})^2+b_2\alpha(1-\Omega_{\rm m})+b_3 \alpha^2\,,
\end{eqnarray}
where $a_1={3(1-w)(1-3w/2)}/[{(5-{6w})^2(5-12w)}]$, $a_2=(31-66w+36w^2)/{(5-6w)^2}$, $b_1=(194-1325w+3039w^2-2880w^3+972w^4)/[4(5-6w)^3(5-12w)(5-18w)]$, $b_2=(-8465+66042w-201168w^2 + 29916w^3-217728w^4 + 62208w^5)/[2(6w-5)^3(5-12w)^2]$, and $b_3=12(1-w)/(5-6w)^3$.

It can be observed that the growth index $\gamma(z)$ recovers to the parameterized form in Eq.~\eqref{gamma-3} within the $w$CDM model, when $\alpha=0$. The relative discrepancy between the parameterization $\Omega_{\rm m}^{\gamma(z)}$ and the growth rate $f$ can be derived by numerically solving Eq.~\eqref{wcdmfeq2} for both a constant $\gamma$ and the second-order approximations of $\gamma(z)$. The corresponding results are shown in the left and right panel of Fig.~\ref{fig:error_wcdm}, respectively. The deviation between the parameterization $\Omega_{\rm m}^\gamma$ and the growth rate $f$ occurs in the relatively low redshift range (particularly $0<z<0.5$), and this discrepancy increases with the increase of the absolute value of the coupling constant $\alpha$. This occurs because the parameterization is derived by expanding Eq.~\eqref{gamma-4} around $\Omega_{\rm m}\sim1$, which holds only for $z\geq1$. In the case of the constant $\gamma$ as seen from the left panel, it is evident that the greatest relative discrepancy between the parameterization $\Omega_{\rm m}^\gamma$ and the growth rate $f$ is about 12\% for $\alpha=0.02$. This level of discrepancy is comparable to the typical observational uncertainty on $f\sigma_8$ ($\sim 15\%$)~\cite{Howlett2017,Avila2021,Bautista2021,Gil2020}. Thus, this discrepancy leads to bias when attempting to constrain the parameter using observations pertinent to the growth rate $f$ in the context of the interacting $w$CDM model. Therefore, it becomes imperative to account for the impacts of the interactions between the dark sectors on the growth index $\gamma$.

It can be concluded that the discrepancy between the parametric form $\Omega_{\rm m}^{\gamma(z)}$ and the numerical solution is significantly reduced. The maximum relative difference decreased from $\sim 12\%$ to $\sim 5\%$, achieving an approximately 60\% reduction in error compared to the results obtained when $\gamma$ is treated as a constant in the parameterization $\Omega_{\rm m}^\gamma$. This improved parameterization makes $\Omega_{\rm m}^{\gamma(z)}$ both an accurate and computationally efficient tool for interacting models. It avoids the computational cost and complexity of repeatedly solving the growth rate differential Eq.~\eqref{wcdmfeq2} numerically, thereby facilitating rapid theoretical predictions in large-scale MCMC analyses.
\begin{figure}[htbp]
	\centering
	\includegraphics[width=0.48\textwidth]{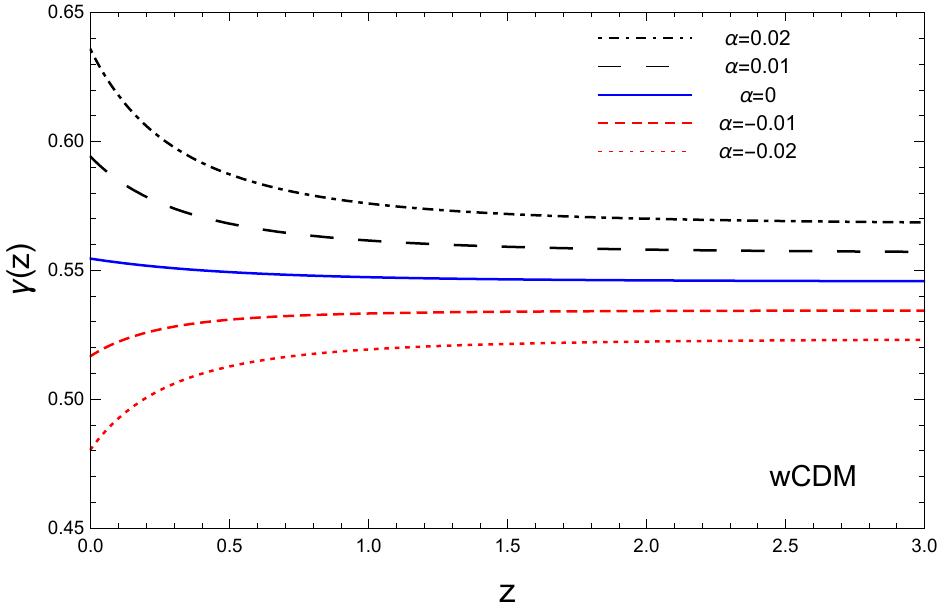}
	\includegraphics[width=0.48\textwidth]{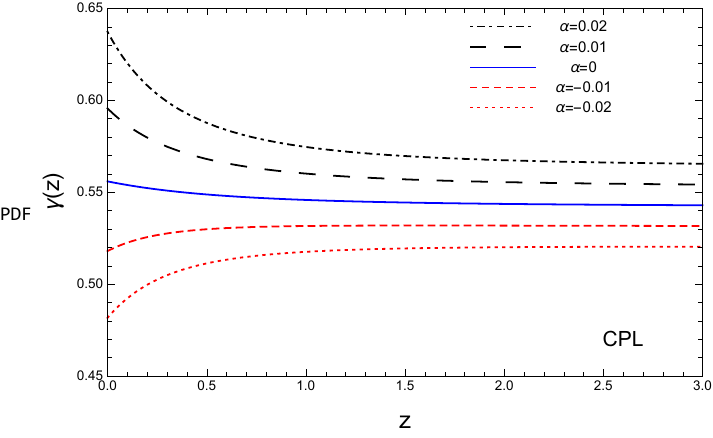}
	\caption{The numerical results of the growth index $\gamma(z)$ as a function of redshift $z$. Left panel (a): the interacting $w$CDM model ($w=-1$). Right panel (b): the interacting CPL model ($w_0=-0.9, w_a=-0.28$).}
	\label{fig:gamma_evolution}
\end{figure}
From our analysis, the coefficient of the first-order correction term $a_2\alpha$ is found to be $a_2\simeq 1.1$ for $w\simeq -1$~\cite{Abdul-Karim2025}. This linear dependence implies that the coupling $\alpha$ induces a shift in the growth index of $\Delta\gamma \simeq 1.1\alpha$. The numerical evaluation of $\gamma(z)$ via Eq.~\eqref{gamma-4}, shown in the left panel of  Fig.~\ref{fig:gamma_evolution}, confirms that the growth index asymptotes to a constant value at relatively high redshifts, while its redshift evolution is predominantly governed by the coupling strength $\alpha$, yielding a nearly uniform shift of approximately $1.1\alpha$  in the $w$CDM scenarios. This behavior is in full agreement with our theoretical predictions.

For the dynamical CPL model, a rigorous analytical Taylor expansion analogous to that in the 
$w$CDM case is precluded by the time-varying equation of state. The numerical evaluation of 
$\gamma(z)$ via Eq.~\eqref{wcdmfeq2} is shown in the right panel of Fig.~\ref{fig:gamma_evolution}. It is observed that the growth index approaches a constant value of 
$0.542$ at relatively high redshifts, while the coupling strength 
$\alpha$ induces an approximately constant shift in the growth index. This shift asymptotically tends to $1.1\alpha$.

Consequently, in both the interacting $w$CDM and CPL frameworks, an interacting dark-energy model with $\alpha>0$ produces a larger growth index, a feature that aligns it phenomenologically with specific modified gravity frameworks, for instance the DGP braneworld scenario ($\gamma_{\rm DGP} \simeq 0.687$) or $f(R)$ models such as the Hu-Sawicki one ($\gamma \gtrsim 0.75$) \cite{Basilakos2017}. On the other hand, a negative coupling $\alpha<0$ leads to a lower value of $\gamma$, mimicking the characteristic behavior of certain viable $f(R)$ theories, e.g., the Starobinsky model ($\gamma \approx 0.4$) \cite{Gannouji2009,starobinsky2007}. This leads to a degeneracy: if future observations find $\gamma\neq 0.55$, growth data alone cannot determine whether the cause is an interaction or modified gravity. In other words, the conventional diagnostic power of the growth index to discriminate between dark energy and modified gravity breaks down when the interaction between the dark sectors is permitted.

To break this degeneracy and determine the allowable range of the coupling strength $\alpha$ and the dark energy equation of state parameters (and thus the permissible shift $\Delta\gamma$), it is essential to confront these interacting models with the latest multi-probe cosmological observations. The subsequent sections present such an analysis, combining background and growth data to place stringent constraints on the dark-sector dynamics, thereby quantifying the extent to which the interaction-induced correction can mimic modified gravity effects and clarifying the interpretative framework for future high-precision growth measurements.


\section{Data and Methodology}
Here, we review the fundamental equations and statistical methodologies employed in observational measurements of cosmic backgrounds, specifically focusing on SNIa, BAO, Cosmic Microwave Background (CMB), and Hubble ($H(z)$) data. Additionally, we incorporate growth rate data in the form of $f\sigma_8$ derived from redshift-space distortion (RSD) observations to constrain the free parameters of both the interacting $w$CDM and dynamical CPL models.

\subsection{SNIa data}
SNIa observations are among the most important probes of cosmic dynamics due to their standardizable nature, and they continue to provide some of the strongest constraints on dark energy models. Here, we employ the Pantheon+ compilation sample of SNIa data, which includes 1701 SNe Ia light curves observed from 1550 distinct SNe and covers the redshift range $0.001<z<2.26$~\cite{Scolnic2022,Brout2022}. The luminosity distances of the Pantheon+ compilation are calibrated from the SALT2 light-curve fitter through applying the Bayesian Estimation Applied to Multiple Species with Bias Corrections method to determine the nuisance parameters~\cite{Scolnic2022}. The distance modulus $\mu(z)$ of any SNIa located at a redshift $z$ is given as $\mu(z)= m_{\rm B}-M_{\rm B}$. The chi-square of the SNIa measurements is given by
\begin{equation}
	\chi^2_{\rm SN}=(\mu_{{\rm obs}}-\mu_{{\rm th}})C^{-1}_{\rm SN}(\mu_{{\rm obs}}-\mu_{{\rm th}})^{\rm T}\,,
\end{equation}
where $\mu_{\rm obs}$ is the observed distance modulus and $C_{\rm SN}$ is the full covariance matrix including both statistical and systematic uncertainties~\cite{Brout2022}.

\subsection{$H(z)$ data}
In our analysis, we employ a widely referenced compilation of Hubble $H(z)$ data (see Ref.~\cite{Cai2022}), which contains 32 data points of $H(z)$ from Refs.~\cite{Simon2005,Stern2010,Moresco2012,Zhang2014,Avila2022,Moresco2016,Ratsimbazafy2017,Borghi2022}, spanning the redshift range $0.07\leq z\leq1.965$. The $\chi^2_{H}$ statistic is calculated using the total covariance matrix, which is decomposed into statistical and systematic components. The systematic part further includes uncertainties from stellar metallicity, residual young components in galaxy spectra, and modeling uncertainties related to star formation history, initial mass functions, and stellar population synthesis models. Although the combination of CMB, BAO, and SNIa data dominates the background constraints, we retain the cosmic chronometer data in our joint analysis. Unlike SNIa and BAO, which provide integrated distance measurements, cosmic chronometers offer a direct, unintegrated probe of the cosmic expansion history. Therefore, while the inclusion of $H(z)$ data does not significantly alter the final parameter contours, it serves as a completely independent astrophysical probe that enhances the diversity and robustness of our multi-probe dataset against potential systematics.

\subsection{CMB data}
The CMB radiation provides crucial insights into the evolution of the early universe. A full analysis of the CMB power spectrum is computationally intensive for non-minimal models~\cite{Zhai2020}. Therefore, for computational efficiency, we adopt the compressed distance prior from the Planck 2018 data release~\cite{Chen2019}, which serves as a reliable substitute for background-level analysis~\cite{Zhai2019}. This prior is characterized by the acoustic scale $l_{\rm A}$, and the shift parameter $R$, which are derived from the angular diameter distance and the comoving sound horizon at the photon decoupling epoch~\cite{Hu1996}. The $\chi^2_{\rm CMB}$ is constructed using these parameters, along with the baryon density $\omega_{\rm b0}$, and their full covariance matrix.

\subsection{BAO data}
BAO measurements are acoustic oscillation patterns in the matter density that serve as a standard ruler for cosmological distances, calibrated by the sound horizon at the drag epoch $r_{\rm d}$~\cite{Aghanim2020}. In this work, we utilize two different sets of BAO data for a comparative analysis.
\subsubsection{SDSS BAO data}
For our first analysis, we use a comprehensive compilation of BAO measurements from all four generations of the Sloan Digital Sky Survey (SDSS). This includes data from SDSS DR7 MGS~\cite{RossAshley2015}, SDSS-III BOSS DR12~\cite{AlamShadab2017}, and the final SDSS-IV eBOSS DR16, which is divided into Luminous Red Galaxies (LRG)~\cite{Bautista2021,Gil2020}, Emission Line Galaxies (ELG)~\cite{De2021,Tamone2020}, the Quasar Sample (QSO)~\cite{Neveux2020,Hou2021}, and the Lyman-$\alpha$ (Ly$\alpha$) forest~\cite{Des2020}. The analysis is performed using the full covariance matrix as provided in Ref.~\cite{Alam2021}.
\subsubsection{DESI DR2 BAO data}
For our second analysis, we replace the SDSS data with the more precise and up-to-date BAO measurements from the DESI DR2~\cite{Abdul-Karim2025}. These data are derived from several distinct tracer samples, including the Bright Galaxy Sample (BGS), LRG, ELGs, QSOs, and the Ly$\alpha$ forest. The specific methodology for utilizing these data and constructing the total $\chi^2$ is detailed in Ref.~\cite{Shen2025}.

\subsection{$f\sigma_8$ data}
The growth rate $f$ is derived from observational data and describes the evolution of matter density fluctuations over time. The observed growth rates can be obtained by various methods, one of which commonly involves large-scale astronomical surveys, such as RSD. However, it is sensitive to the bias parameter $b$, which typically lies in the range of $1-3$. This sensitivity makes the observational $f_{\rm obs}$ data unreliable~\cite{Nesseris2017}. On the other hand, combining $f$ with $\sigma_8$ allows for independence from biases, thereby enhancing the reliability of the observational data. The observational growth rate in the form of $f\sigma_{8}$ data can be obtained from RSD measurements~\cite{Nesseris2017,Huang2022,Alam2016,Pinho2018,Howlett2017}. The expression of $\delta(z)/\delta_0$ is given by
\begin{equation}
	{\delta(z)\over \delta_0}=\exp(-\int_0^z{fd\tilde{z}\over 1+\tilde{z}})\,.
\end{equation}
The complete compilation of the observed growth rate data, $f\sigma_8(z)$, used in our analysis is explicitly provided in Appendix~\ref{app:fs8_data}. These data sets are primarily taken from the compilation presented in Table 2 of Ref.~\cite{Huang2022} (which includes data from~\cite{Bautista2021,Gil2020,Howlett2017,Avila2021,AlamShadab2017,De2021,Tamone2020,Neveux2020,Hou2021,Huterer2017,Turnbull2012,Hudson2012,Davis2011,Song2009,Simpson2016,Pezzotta2017,Okumura2016,Blake2012,Howlett2015}), including 20 measurements from various surveys. The $\chi^{2}_{f\sigma_8}$ from the independent data points is calculated using the following equation,
\begin{equation}
	\chi^2_{f\sigma_8}=\sum_{i=1}^{11}{[f\sigma_{8,{\rm obs}}(z_i)-f\sigma_{8,{\rm mod}}(z_i)]^2\over\sigma^2_{f\sigma_8}(z_i)}\,.
\end{equation}
To account for the correlations among the three distinct subsets of WiggleZ data, we utilize the covariance matrix in Ref.~\cite{Pinho2018} to compute the chi-squared statistic $\chi^{2}_{\rm WiggleZ}$.

Note also that the compilation from the eBOSS DR16~\cite{Alam2021} re-analyzes all four generations of SDSS data, and then incorporates the systematic errors and consensus estimates into the covariance matrices to obtain the combined BAO+RSD measurements with inclusion of both the Alcock-Paczynski (AP) effect and the reconstruction procedure~\cite{Alcock1979}. In RSD measurements, assuming a fiducial cosmological model when converting redshifts to distances introduces additional anisotropy, known as the AP effect. To reduce the bias introduced by the AP effect, the model estimation of $f\sigma_8(z)$ should be corrected as~\cite{Alam2016,Howlett2017}
\begin{equation}
	f\sigma_8^{\rm corrected}={H^{\rm model}(z)D_{\rm A}^{\rm model}(z)\over H^{\rm fid}(z)D_{\rm A}^{\rm fid}(z)}\times (f\sigma_8)^{\rm model}
\end{equation}
The superscript ``model" refers to the interacting dark energy model used, while the superscript ``$\rm fid$" denotes the fiducial flat $\Lambda$CDM cosmology assumed in RSD measurements.

To constrain the model parameters, we must compare these observations with theoretical predictions. In this work, we compute the theoretical value of $f\sigma_8(z)$ using the two distinct methods developed in Section III:
\begin{itemize}
	\item \textbf{Method I (Numerical)}: The theoretical growth rate $f(z)$ is obtained by numerically solving the modified differential equation (Eq.~\eqref{wcdmfeq2}). This method is exact and is applied to both the interacting $w$CDM and dynamical CPL models.
    \item \textbf{Method II (Parametric)}: We use the computationally efficient approximation $f(z) \approx \Omega_{\rm m}(z)^{\gamma(z)}$, where $\gamma(z)$ is given by the second-order approximation in Eq.~\eqref{gamma-5}. This method is applied exclusively to the interacting $w$CDM model.
\end{itemize}

All the likelihood information pertaining to the completed SDSS-IV is consolidated on the public SDSS svn repository. The overall $\chi^{2}_{\rm RSD}$ statistic for all of the RSD measurements is given by:
\begin{equation}
	\chi^{2}_{\rm {RSD}}=\chi^2_{f\sigma_8}+\chi^{2}_{\rm WiggleZ}+\chi^{2}_{\rm SDSS}\,.
\end{equation}

It is important to note that we use this RSD compilation in two distinct ways depending on the background BAO dataset employed (SDSS or DESI) to strictly avoid double-counting the BAO information. The detailed treatment of the SDSS measurements (whether using the combined BAO+RSD likelihoods or the RSD-only measurements) is explicitly described in Appendix~\ref{app:fs8_data}. This approach allows us to consistently test the constraining power of existing growth data, while acknowledging that a dedicated RSD analysis for DESI DR2 is still forthcoming.

\subsection{Observational Constraints}
In this section, we present constraints on the free parameters of the interacting dark energy models. For the interacting $w$CDM scenario, the parameter space is $\{H_0, \Omega_{\mathrm{m},0}, w, \alpha, \sigma_8\}$. For the dynamical CPL scenario, the parameter space is extended to $\{H_0, \Omega_{\mathrm{m},0}, w_0, w_a, \alpha, \sigma_8\}$. Our analysis is structured around two primary datasets, each combining background cosmological probes:
\begin{itemize}
	\item Background dataset with SDSS BAO (denoted as $\mathrm{SHCB}_{\mathrm{SDSS}}$): SNIa + $H(z)$ + CMB + SDSS BAO.
	\item Background dataset with DESI BAO (denoted as $\mathrm{SHCB}_{\mathrm{DESI}}$): SNIa + $H(z)$ + CMB + DESI BAO.
\end{itemize}
For each dataset, we first perform an analysis using background data alone, and then extend it by incorporating RSD measurements. For cases including RSD data, we compute the theoretical growth rate $f(z)$ using the two independent methodologies (Method I and Method II) described above. This dual-methodology approach allows us to rigorously validate the accuracy of our parametric approximation against the exact numerical solution for the $w$CDM case, and to assess its computational advantages in MCMC analyses.

The total $\chi^2$ functions for our analyses are defined accordingly for each dataset combination. For background-only analyses, we compute:
\begin{equation}
	\chi^2_{\rm{SHCB}} = \chi^2_{\rm{SN}} + \chi^2_{H(z)} + \chi^2_{\rm{CMB}} + \chi^2_{\rm{BAO}}\,,
\end{equation}
where $\chi^2_{\rm{BAO}}$ represents either the SDSS or DESI BAO measurements as specified.

When incorporating RSD data, we consider Method I and Method II for calculating the theoretical growth rate $f(z)$. The corresponding total $\chi^2$ functions are:
\begin{itemize}
	\item For joint analysis using Method I:
	\begin{equation}
		\chi^2_{\rm{tot,\,I}} = \chi^2_{\rm{SHCB}} + \chi^2_{{\rm RSD},\,{\rm I}}\,.
	\end{equation}
	\item For joint analysis using Method II:
	\begin{equation}
		\chi^2_{\rm{tot,\,II}} = \chi^2_{\rm{SHCB}} + \chi^2_{{\rm RSD},\,{\rm II}}\,.
	\end{equation}
\end{itemize}
Here, $\chi^2_{{\rm RSD},\,{\rm I}}$ and $\chi^2_{{\rm RSD},\,{\rm II}}$ denote the chi-square statistics derived from the RSD data. They are computed by comparing the observed $f\sigma_8$ measurements with the corresponding theoretical predictions for the growth rate $f(z)$ obtained using Method I and Method II, respectively.

In this study, we use the MCMC package CosmoMC to determine the posterior distributions for these analyses~\cite{Lewis2002,Lewis2013}. We assess the convergence of the MCMC chains using the Gelman-Rubin statistic, requiring $R-1\leq 0.01$~\cite{Gelman1992}. The analysis of the MCMC chains is performed using the public package GetDist~\cite{Lewis2019}.

\begin{figure}[htbp]
	\centering
	\includegraphics[width=0.9\textwidth]{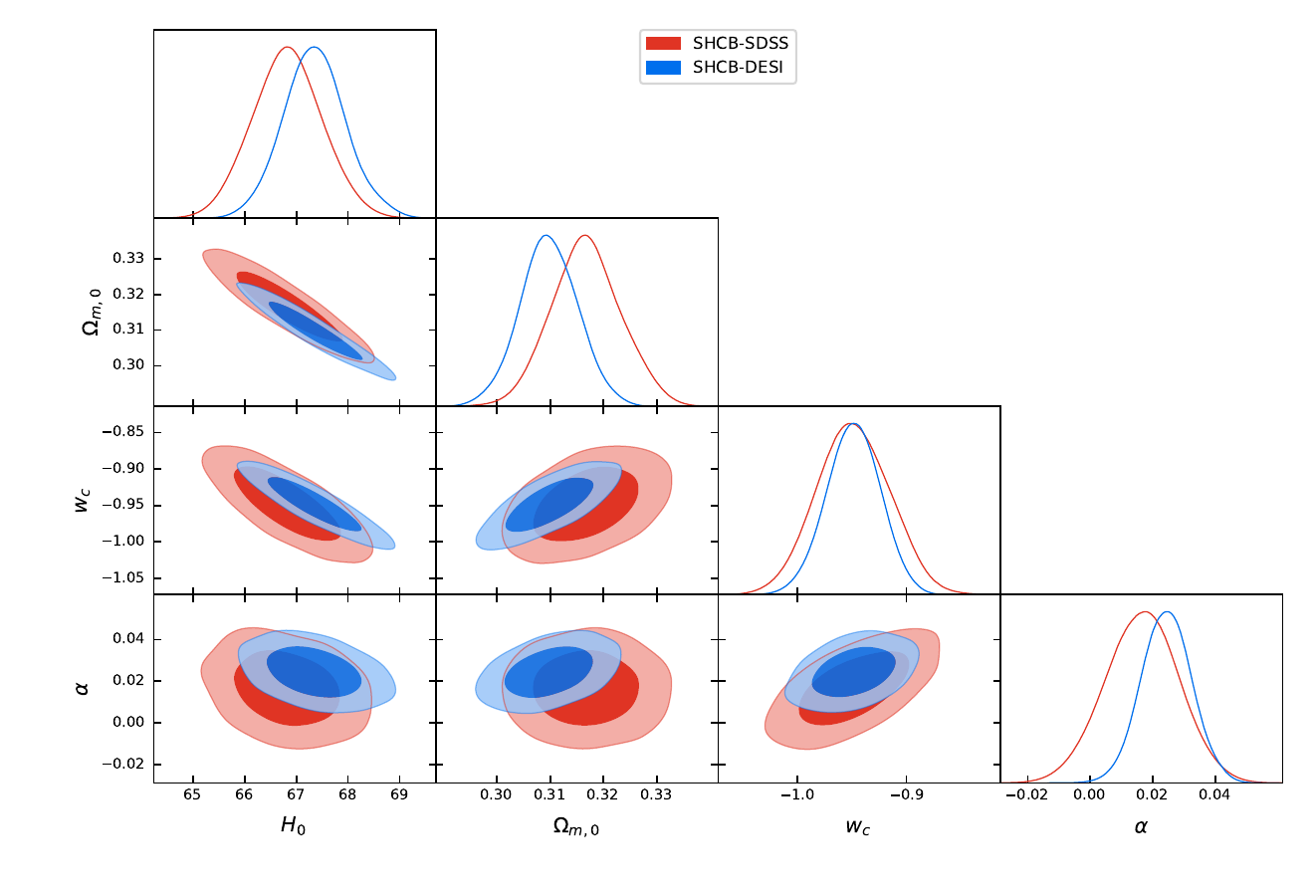}
	\caption{Triangular plot showing the 2D joint and 1D marginalized posterior probability distributions for the cosmological parameters of the interacting $w$CDM model, constrained by the background-only datasets: SHCB$_{\rm SDSS}$ (red contours) and SHCB$_{\rm DESI}$ (blue contours).}
	\label{fig:mcmc_wcdm_bg}
\end{figure}

\begin{figure}[htbp]
	\centering
	\includegraphics[width=0.9\textwidth]{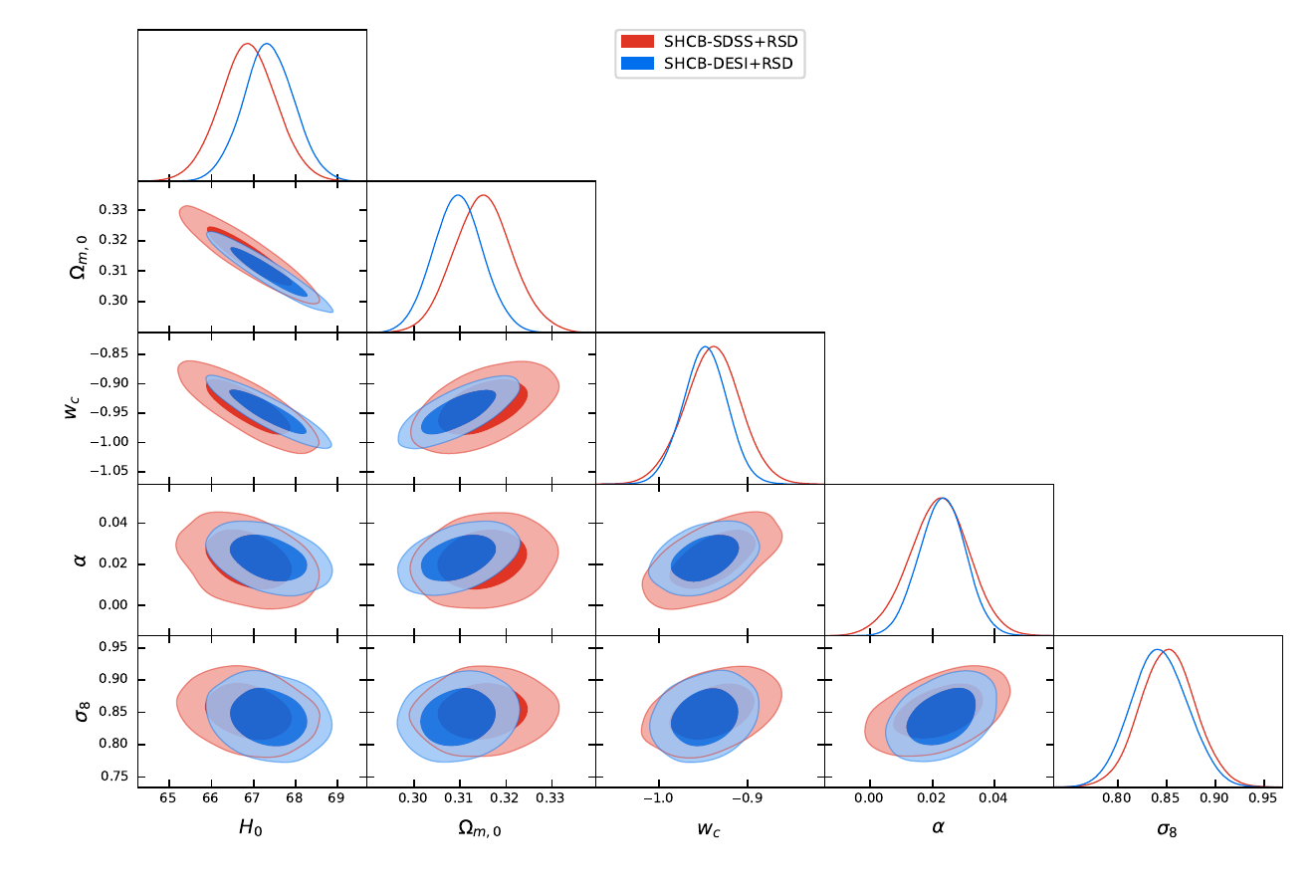}
	\caption{Triangular plot for the interacting $w$CDM model constrained by the joint background and RSD datasets: SHCB$_{\rm SDSS}$+RSD (red contours) and SHCB$_{\rm DESI}$+RSD (blue contours). The theoretical growth rate $f(z)$ is computed via direct numerical integration (Method I).}
	\label{fig:mcmc_wcdm_rsd1}
\end{figure}

\begin{figure}[htbp]
	\centering
	\includegraphics[width=0.9\textwidth]{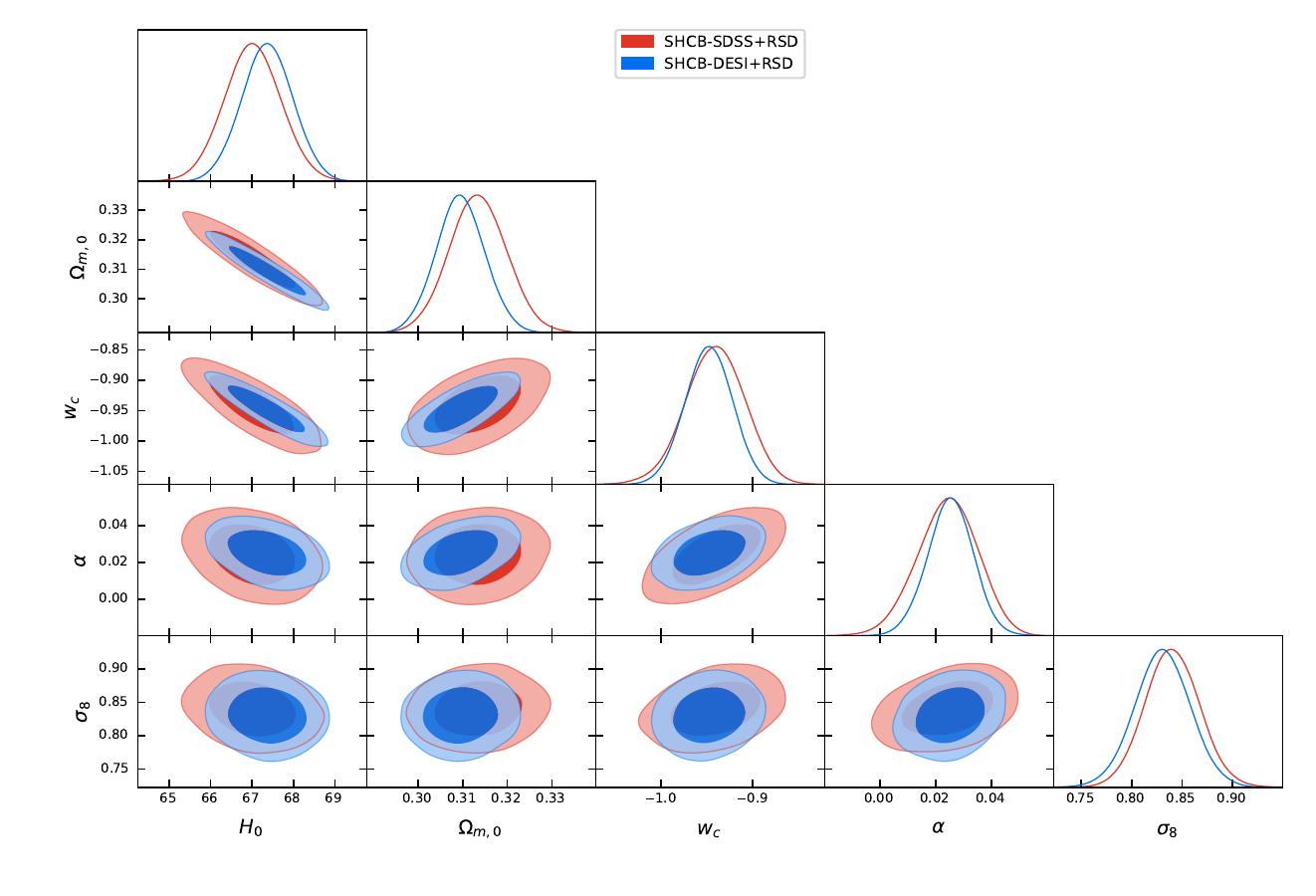}
	\caption{Triangular plot for the interacting $w$CDM model constrained by the joint background and RSD datasets: SHCB$_{\rm SDSS}$+RSD (red contours) and SHCB$_{\rm DESI}$+RSD (blue contours). The theoretical growth rate $f(z)$ is computed using the improved analytical parameterization (Method II).}
	\label{fig:mcmc_wcdm_rsd2}
\end{figure}

\begin{figure}[htbp]
	\centering
	\includegraphics[width=0.9\textwidth]{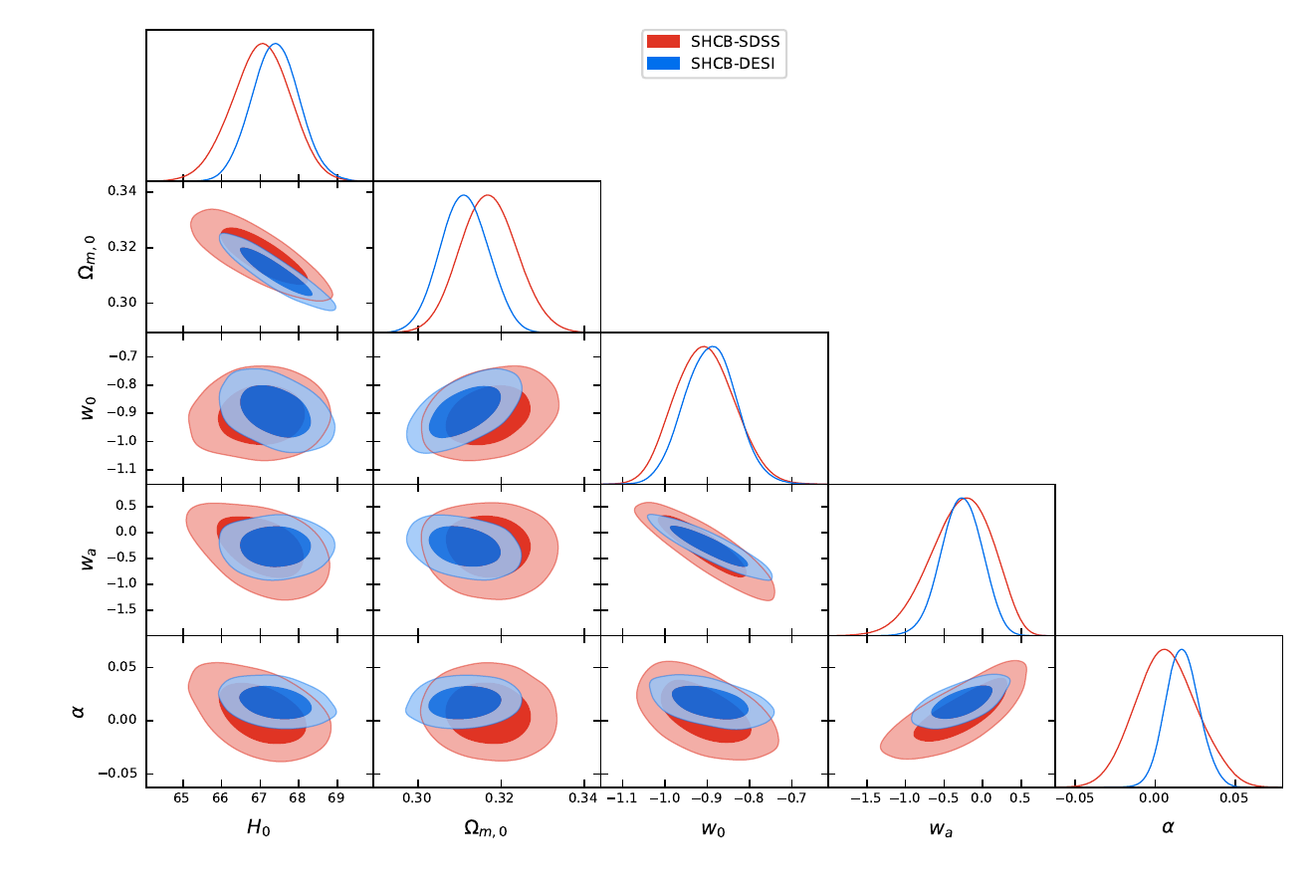}
	\caption{Triangular plot showing the 2D joint and 1D marginalized posterior probability distributions for the cosmological parameters of the interacting CPL model, constrained by the background-only datasets: SHCB$_{\rm SDSS}$ (red contours) and SHCB$_{\rm DESI}$ (blue contours).}
	\label{fig:mcmc_cpl_bg}
\end{figure}

\begin{figure}[htbp]
	\centering
	\includegraphics[width=0.9\textwidth]{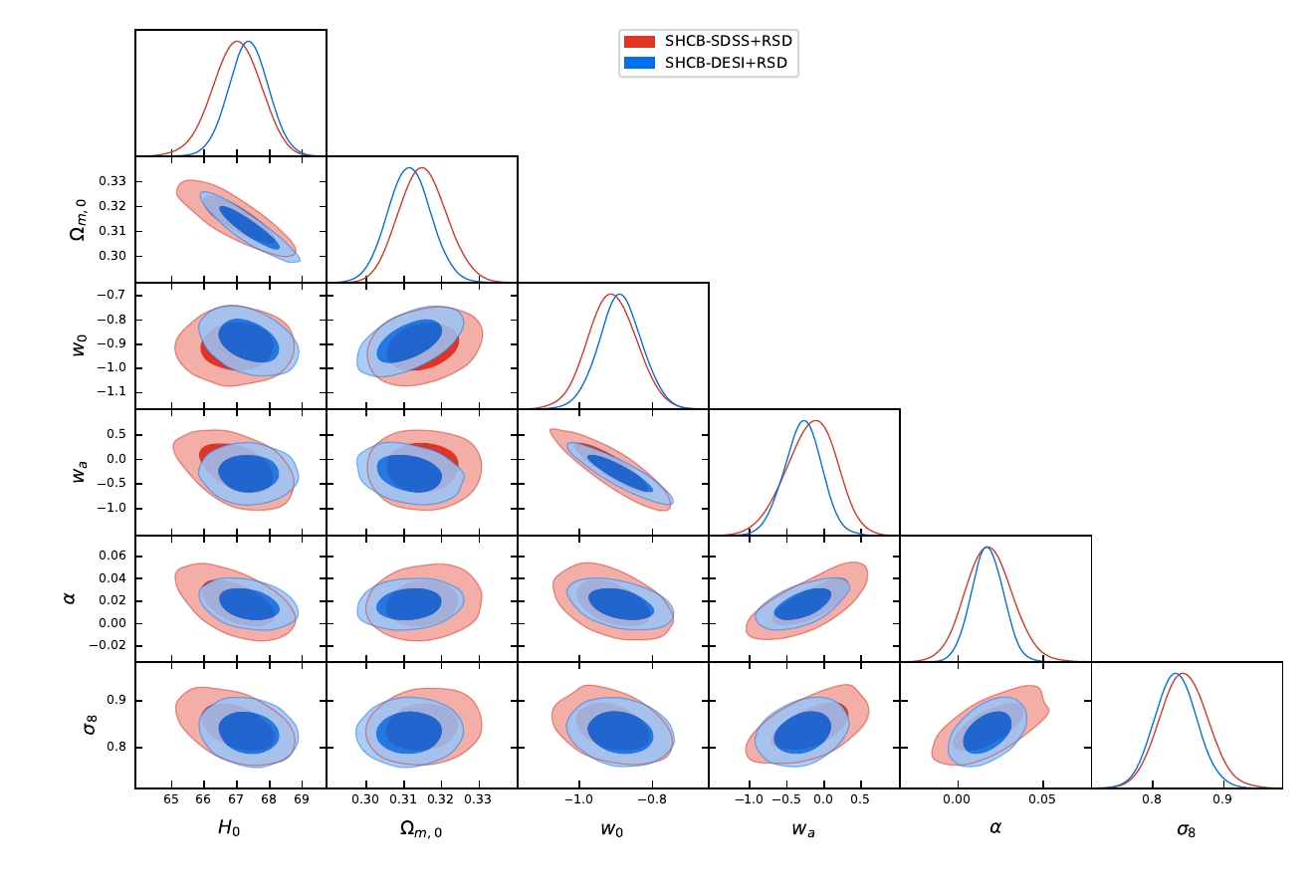}
	\caption{Triangular plot for the interacting CPL model constrained by the joint background and RSD datasets: SHCB$_{\rm SDSS}$+RSD (red contours) and SHCB$_{\rm DESI}$+RSD (blue contours). The theoretical growth rate $f(z)$ is computed via direct numerical integration (Method I).}
	\label{fig:mcmc_cpl_rsd1}
\end{figure}

The marginalized constraints for the interacting $w$CDM model are summarized in Table~\ref{tab:wcdm_sdss} (using SDSS BAO) and Table~\ref{tab:wcdm_desi} (using DESI DR2 BAO). The corresponding constraints for the dynamical CPL model are presented in Table~\ref{tab:cpl_merged}. To provide a comprehensive visual comparison, the posterior distributions for the $w$CDM model are displayed in the triangular plots of Figs.~\ref{fig:mcmc_wcdm_bg}--\ref{fig:mcmc_wcdm_rsd2}. These figures illustrate the two-dimensional (2D) joint and one-dimensional (1D) marginalized posterior probability distributions for the cosmological parameters. The parameter $w_c$ in these figures represents the constant equation of state $w$ of dark energy. Additionally, the corresponding 1D and 2D posterior distributions for the interacting CPL model, showing the results with and without RSD data for both SDSS and DESI combinations, are illustrated in Figs.~\ref{fig:mcmc_cpl_bg} and \ref{fig:mcmc_cpl_rsd1}.


\begin{table}[htp]
	\centering
	\begin{tabular}{c c c c}
		\hline\hline
		\ \ \  Parameter \ & \ \ \ SHCB$_{\rm SDSS}$ \ \ \ \ & \ \ SHCB$_{\rm SDSS}$+${\rm RSD}^{\ast}$ & \ \ SHCB$_{\rm SDSS}$+${\rm RSD}^{\diamond}$ \\
		\hline
		$H_0$ & $66.82\pm 0.68 $ & $66.87 \pm 0.68$ & $67.01\pm 0.69$ \\
		$\Omega_{\rm m,0}$ & $0.3167\pm 0.0065 $ & $0.3150 \pm 0.0064$ & $0.3134\pm0.0065$ \\
		$w$ & $-0.949\pm 0.033$ & $-0.940 \pm 0.032$ & $-0.941\pm 0.032$ \\
		$\alpha$ & $0.017\pm 0.012$ & $0.0221 \pm 0.0095$ & $0.024^{+0.011}_{-0.010}$ \\
		$\sigma_8$ & -- & $0.851 \pm 0.028$ & $0.840\pm 0.027 $ \\
		\hline
	\end{tabular}
	\caption{Constraints on the interacting $w$CDM model from different data combinations using SDSS BAO. Column 2: SHCB$_{\rm SDSS}$-only. Columns 3 and 4: SHCB$_{\rm SDSS}$ + RSD, with the theoretical growth rate $f(z)$ computed via Method I ($\ast$) and Method II ($\diamond$), respectively. $H_0$ is in units of $\rm{km\,s^{-1}\,Mpc^{-1}}$.}
	\label{tab:wcdm_sdss}
\end{table}


\begin{table}[htp]
	\centering
	\begin{tabular}{c c c c}
		\hline\hline
		\ \ \  Parameter \ & \ \ \ SHCB$_{\rm DESI}$ \ \ \ \ & \ \ SHCB$_{\rm DESI}$+${\rm RSD}^{\ast}$ & \ \ SHCB$_{\rm DESI}$+${\rm RSD}^{\diamond}$ \\
		\hline
		$H_0$ & $67.35\pm 0.61 $ & $67.36 \pm 0.61$ & $67.37\pm 0.61$ \\
		$\Omega_{\rm m,0}$ & $0.3098\pm 0.0055 $ & $0.3096 \pm 0.0054$ & $0.3095\pm0.0055$ \\
		$w$ & $-0.949\pm 0.025$ & $-0.948 \pm 0.025$ & $-0.948\pm 0.025$ \\
		$\alpha$ & $0.0243\pm 0.0080$ & $0.0230 \pm 0.0076$ & $0.0252\pm 0.0082$ \\
		$\sigma_8$ & -- & $0.843 \pm 0.029$ & $0.831\pm 0.028 $ \\
		\hline
	\end{tabular}
	\caption{Constraints on the interacting $w$CDM model from different data combinations using DESI DR2 BAO. Column 2: SHCB$_{\rm DESI}$-only. Columns 3 and 4: SHCB$_{\rm DESI}$ + RSD, with the theoretical growth rate $f(z)$ computed via Method I ($\ast$) and Method II ($\diamond$), respectively.}
	\label{tab:wcdm_desi}
\end{table}

\begin{table}[htp]
	\centering
	\begin{tabular}{c c c c c}
		\hline\hline
		\ \ \  Parameter \ & \ \ \ SHCB$_{\rm SDSS}$ \ \ \ \ & \ \ SHCB$_{\rm SDSS}$+${\rm RSD}^{\ast}$ & \ \ \ SHCB$_{\rm DESI}$ \ \ \ \ & \ \ SHCB$_{\rm DESI}$+${\rm RSD}^{\ast}$ \\
		\hline
		$H_0$ & $67.02\pm 0.75$ & $66.99\pm 0.74$ & $67.40\pm 0.61$ & $67.36\pm 0.62$ \\
		$\Omega_{\rm m,0}$ & $0.3170\pm 0.0068$ & $0.3151\pm 0.0063$ & $0.3112\pm 0.0057$ & $0.3114\pm 0.0058$ \\
		$w_0$ & $-0.906\pm 0.070$ & $-0.910\pm 0.066$ & $-0.893\pm 0.061$ & $-0.888\pm 0.059$ \\
		$w_a$ & $-0.28^{+0.45}_{-0.35}$ & $-0.18^{+0.37}_{-0.30}$ & $-0.27\pm 0.26$ & $-0.28\pm 0.25$ \\
		$\alpha$ & $0.007\pm 0.019$ & $0.018\pm 0.014$ & $0.017\pm 0.010$ & $0.0173\pm 0.0095$ \\
		$\sigma_8$ & -- & $0.844\pm 0.034$ & -- & $0.833\pm 0.030$ \\
		\hline
	\end{tabular}
	\caption{Constraints on the interacting CPL model ($w(a) = w_0 + w_a(1-a)$) from different data combinations using SDSS BAO and DESI DR2 BAO. The theoretical growth rate $f(z)$ for the RSD datasets is computed via Method I ($\ast$).}
	\label{tab:cpl_merged}
\end{table}
\clearpage

\section{Analysis and Results}

First, the background-only SHCB$_{\rm SDSS}$ dataset constrains the coupling constant to $\alpha = 0.017 \pm 0.012$ and the equation of state to $w = -0.949 \pm 0.033$. Using our improved parameterization (Method II), we find $\alpha = 0.024^{+0.011}_{-0.010}$ and $w = -0.941 \pm 0.032$ for the SHCB$_{\rm SDSS}$+RSD combination. This represents a $\sim 12\%$ improvement in precision (with $\delta\alpha$ reduced from $\sim 0.012$ to $\sim 0.0105$) compared to the background-only case. Our constraint is approximately 85\% more precise than the value $\alpha = -0.197 \pm 0.071$ reported in Ref.~\cite{Li2024}, which used DESI BAO and SNIa data. Similarly, for the matter clustering amplitude, we obtain $\sigma_8 = 0.840 \pm 0.027$ from the SHCB$_{\rm SDSS}$+RSD analysis. This uncertainty is about 80\% smaller than that reported in Ref.~\cite{Nazari Pooya2024} using a similar combination of background and growth data. By rigorously including the radiation component at high redshifts and combining it with the latest RSD data, our analysis finds that the coupling $\alpha$ is consistent with zero at approximately the $2.3\sigma$ confidence level, and $w$ is consistent with $-1$ at approximately the $1.9\sigma$ confidence level.

Next, we analyze the constraints derived from the DESI-based datasets (Table~\ref{tab:wcdm_desi}), which provide a remarkable improvement in precision. The background-only SHCB$_{\rm DESI}$ dataset yields a tight constraint of $\alpha = 0.0243 \pm 0.0080$. When RSD data are included (SHCB$_{\rm DESI}$+RSD, Method I), we obtain $\alpha = 0.0230 \pm 0.0076$ and $w = -0.948 \pm 0.025$. This indicates that for the $w$CDM model, $\alpha$ is consistent with zero at approximately the $3\sigma$ confidence level and $w$ is consistent with $-1$ at the $2\sigma$ level. While the central values suggest a positive $\alpha$ (implying energy transfer from dark energy to dark matter) and $w > -1$, the results remain statistically compatible with the standard $\Lambda$CDM paradigm. Nevertheless, the tightened constraints highlight the exceptional constraining power of the new DESI DR2 BAO measurements when rigorously combined with the radiation component at high redshifts.

To further explore the dynamical nature of dark energy, we extend our analysis to the interacting CPL model, with results summarized in Table~\ref{tab:cpl_merged}. For the SHCB$_{\rm DESI}$+RSD dataset using exact numerical integration (Method I), we find $w_0 = -0.888 \pm 0.059$ and $w_a = -0.28 \pm 0.25$, alongside a coupling constant $\alpha = 0.0173 \pm 0.0095$. The corresponding analysis using the SHCB$_{\rm SDSS}$+RSD dataset yields similar trends ($w_0 = -0.910 \pm 0.066$, $\alpha = 0.018 \pm 0.014$). These results demonstrate that for the CPL model, $\alpha$, $w_0$, and $w_a$ are all consistent with their $\Lambda$CDM values ($0$, $-1$, and $0$) within the $2\sigma$ confidence level. Overall, the results show no definitive statistical evidence for a departure from the standard $\Lambda$CDM cosmology.

We perform a consistency check by comparing the constraints derived from our two independent methods for computing the growth rate. For the $w$CDM model with the SHCB$_{\rm DESI}$+RSD combination, the direct numerical integration (Method I) yields $\alpha = 0.0230 \pm 0.0076$, while the improved parameterization (Method II) gives $\alpha = 0.0252 \pm 0.0082$. Across both datasets for the $w$CDM model, the central values and $1\sigma$ uncertainties obtained via the two methods show robust consistency. This serves as a strong validation of our analytical second-order approximation for the growth index $\gamma(z)$ in Eq.~\eqref{gamma-5}. The parameterized approach provides a substantial gain in computational efficiency by avoiding repeated numerical integration across the high-dimensional parameter space, making it a highly practical tool for cosmological MCMC analyses.

Comparing the constraints on the common parameters between the interacting $w$CDM and dynamical CPL models reveals several notable trends. First, the background cosmological parameters ($H_0$ and $\Omega_{\rm m,0}$) as well as the matter clustering amplitude ($\sigma_8$) remain highly robust; their central values and uncertainties are nearly identical across both models. However, the introduction of the dynamical parameter $w_a$ in the CPL model inevitably introduces parameter degeneracies, which broadens the constraints on the dark sector interaction. Specifically, the constraint on the coupling constant $\alpha$ becomes slightly looser in the CPL framework, with the $1\sigma$ uncertainty increasing from $\sim 0.0076$ to $\sim 0.0095$. Interestingly, the best-fit value of $\alpha$ shifts closer to zero in the CPL model (from $0.0230$ to $0.0173$). This shift suggests that allowing for a time-varying dark energy equation of state can partially absorb the phenomenological signatures that would otherwise be attributed to the dark-sector interaction, thereby reducing the statistical preference for a non-zero coupling.

From the two-dimensional posterior distributions presented in Figs.~\ref{fig:mcmc_wcdm_bg}--\ref{fig:mcmc_wcdm_rsd2}, a clear positive correlation can be observed between the dark energy equation of state $w$ and the interaction coupling $\alpha$ in the $w$CDM model. This indicates a physical degeneracy: a more positive coupling (stronger energy transfer to dark matter) prefers a less negative equation of state to maintain the observed expansion history. Interestingly, in the dynamical CPL model (Figs.~\ref{fig:mcmc_cpl_bg} and \ref{fig:mcmc_cpl_rsd1}), this degeneracy becomes more complex due to the time-varying nature of the equation of state. We observe a negative correlation between $w_0$ and $\alpha$, but a positive correlation between $w_a$ and $\alpha$. These correlations emphasize that, in interacting dark energy models, both the background dynamics and the coupling must be fitted simultaneously; fixing any of these parameters to their $\Lambda$CDM values could severely bias the estimates of the others and mask the true underlying mechanism.

As noted in Section III, the coupling $\alpha$ introduces a first-order correction to the growth index, $\Delta\gamma \simeq 1.1\alpha$, in both models. This term is responsible for the theoretical degeneracy, allowing an IDE model to mimic the growth index signature of certain modified gravity theories. The observational constraints can be used to directly quantify the magnitude of this effect. Taking the SHCB$_{\rm DESI}$+RSD constraint for the $w$CDM model ($\alpha = 0.0230 \pm 0.0076$), the $3\sigma$ upper bound on the coupling restricts the growth index to the interval $0.5469 \lesssim \gamma \lesssim 0.5974$. For the dynamical CPL model, using the corresponding $3\sigma$ interval for the coupling ($\alpha = 0.0173 \pm 0.0095$) alongside the numerical results displayed in the right panel of Fig.~\ref{fig:gamma_evolution}, we find that the growth index is constrained to approximately $0.53 \lesssim \gamma \lesssim 0.59$ at the $3\sigma$ CL. Therefore, the observational constraints strongly disfavor the region of parameter space where interacting dark energy can mimic modified gravity, restricting the growth index to a common approximate interval of $0.53 \lesssim \gamma \lesssim 0.60$ for both models. This range is too narrow to span the gap between the $\Lambda$CDM value ($\gamma \approx 0.55$) and the predictions of representative modified gravity theories, such as DGP ($\gamma_{\mathrm{DGP}} \simeq 0.687$) or some viable $f(R)$ models ($\gamma \sim 0.40\text{--}0.43$). Consequently, while the degeneracy between IDE and modified gravity remains at the theoretical level, the present observational bounds strongly constrain the parameter space where IDE could mimic modified gravity, making the two scenarios distinguishable with current data.

Looking forward, this conclusion refines the interpretative framework for future high-precision growth-of-structure measurements from surveys like DESI (full RSD), Euclid, and the Square Kilometre Array (SKA). Should such experiments detect a statistically significant deviation of the growth index $\gamma$ from the $\Lambda$CDM value of $\sim 0.55$, our analysis implies that the explanation would be more strongly inclined towards a genuine modification of gravity on cosmological scales, rather than a non-minimal interaction between dark matter and dark energy. Therefore, our work not only places stringent constraints on possible interactions within the dark sector, but also establishes the cosmological diagnostic tool for distinguishing between IDE and modified gravity.


\section{Conclusion}\label{sec6}
IDE models offer a natural mechanism for an effective dynamical dark energy equation of state through energy transfer between dark matter and dark energy \cite{Guedezounme2025,Westhuizen2025a,Westhuizen2025b}. This theoretical appeal has been further highlighted by recent observational analyses. For instance, studies combining the Dark Energy Spectroscopic Instrument (DESI) DR2 \cite{Abdul-Karim2025} with the re-calibrated DES SNIa sample \cite{Popovic2025} have reported a statistical preference for a time-varying dark energy equation of state over a cosmological constant ($\Lambda$), renewing interest in IDE as a potential framework to accommodate such dynamics. However, whether such an interaction is supported by observations remains an open question, necessitating rigorous tests against a broad array of cosmological probes.

In this work, we have investigated the impact of a non-gravitational coupling between dark matter and dark energy on the growth of linear matter density perturbations. To ensure a comprehensive and rigorous analysis, we considered both the interacting $w$CDM model and the dynamical CPL parameterization, and explicitly incorporated the radiation component to guarantee the accuracy of high-redshift background evolution for CMB data. We demonstrated that such an interaction directly alters the growth rate $f$. For the $w$CDM model, we derived a second-order approximation for the growth index $\gamma(z)$. This approach leads to an improved and computationally efficient $\Omega_{\rm m}^{\gamma(z)}$ parameterization for the $w$CDM model, whose accuracy was validated against exact numerical integration, restricting the maximum relative error to approximately $5\%$. For the CPL model, the numerical results for the growth index are also presented. A key theoretical insight emerging from our analysis is that the coupling introduces a universal correction to the growth index, approximately $1.1\alpha$ in both models. This correction implies a potential degeneracy: an IDE model can mimic the growth-index signature of certain modified-gravity theories (e.g., DGP or $f(R)$ gravity). Consequently, a measurement of $\gamma$ alone cannot distinguish between a non-minimal coupling in the dark sector and a genuine modification of gravity.

We confronted the models with the latest multi-probe cosmological data. Using background observations (Pantheon+ SNIa, $H(z)$, CMB, and SDSS/DESI BAO) together with redshift-space distortion measurements, we obtained tight constraints on the dark-sector dynamics. The DESI-based analysis for the $w$CDM model gives $\alpha = 0.0230 \pm 0.0076$, consistent with zero at approximately the $3\sigma$ confidence level, and $w = -0.948 \pm 0.025$, consistent with $-1$ at the $2\sigma$ level. For the dynamical CPL model, we found $\alpha = 0.0173 \pm 0.0095$, with $w_0 = -0.888 \pm 0.059$ and $w_a = -0.28 \pm 0.25$, demonstrating that all parameters are consistent with their $\Lambda$CDM values within the $2\sigma$ confidence level. These results indicate that the non-interacting $\Lambda$CDM paradigm remains in agreement with current observations. Notably, comparing the two models reveals that allowing for a time-varying equation of state partially absorbs the phenomenological signatures of the dark-sector interaction, shifting the best-fit coupling $\alpha$ closer to zero and slightly broadening its uncertainty. Furthermore, our analysis reveals complex correlations between the equation of state parameters and the coupling $\alpha$, which indicates that the dynamical nature of dark energy and the dark matter--dark energy interaction are observationally linked. Therefore, when interpreting deviations from the $\Lambda$CDM paradigm in the future, it is necessary to constrain these parameters jointly in order to more accurately discriminate between interacting dark energy and modified gravity theories.

For the $w$CDM model, the excellent agreement between the constraints obtained from the direct numerical integration of the growth equation (Method I) and those from our improved parameterization (Method II) validates the accuracy of the latter. This makes it a practical tool for future large-scale cosmological explorations. The observational bounds on the coupling translate into strict $3\sigma$ intervals for the growth index: $0.5469 \lesssim \gamma \lesssim 0.5974$ for the $w$CDM model and approximately $0.53 \lesssim \gamma \lesssim 0.59$ for the CPL model, restricting the growth index to a common approximate interval of $0.53 \lesssim \gamma \lesssim 0.60$ for both models. This range is too narrow to span the gap between the $\Lambda$CDM value ($\gamma \approx 0.55$) and the predictions of representative modified gravity theories, such as DGP ($\gamma_{\mathrm{DGP}} \simeq 0.687$) or some viable $f(R)$ models ($\gamma \sim 0.40\text{--}0.43$). Consequently, while the degeneracy between IDE and modified gravity remains at the theoretical level, the present observational bounds strongly constrain the parameter space where IDE could mimic modified gravity, making the two scenarios distinguishable with current data.

Our results refine the interpretative framework for upcoming high-precision growth-of-structure measurements from DESI, Euclid, and SKA. Should next-generation surveys detect a significant deviation of $\gamma$ from the $\Lambda$CDM expectation, the present analysis indicates that the cause would more definitively point towards a genuine modification of gravity on cosmological scales, rather than a non-minimal interaction within the dark sector. Thus, the growth index, especially when combined with precise background data, serves as a robust diagnostic for discriminating between IDE and modified gravity scenarios.
In future work, we will further investigate the characteristics of density perturbations 
in models with other interaction forms.

\section*{Acknowledgments}
We are grateful to Professor Yungui Gong for his helpful guidance over the years in the field of linear density perturbations.
This work was supported by the National Natural Science Foundation of China under Grants No. 12375045,
No. 12305056, No. 12105097, No.12405054, No.12405053, and No. 12205093, the Cultivation Project for Young and Middle aged Teachers in Provincial Colleges and Universities under grant No. YQZD2024034.

\appendix
\renewcommand{\theequation}{\thesection\arabic{equation}}
\setcounter{equation}{0}
\section{Detailed Derivation of the Background Evolution Equations with Radiation}
\label{app:derivation}

In this appendix, we provide the detailed step-by-step derivations for the background evolution equations, specifically focusing on the inclusion of the radiation component ($\rho_{\rm r}$) and the dynamical CPL equation of state, $w(a) = w_0 + w_a(1-a)$. The constant $w$CDM model can be naturally recovered by setting $w_a = 0$ and $w_0 = w$.

\subsection{Evolution of Energy Densities}
Our derivation is based on a flat universe model containing radiation ($\rho_{\rm r}$), matter ($\rho_{\rm m}$), and interacting dark energy ($\rho_{\rm d}$). The continuity equations are:
\begin{eqnarray}
	&& \dot{\rho}_{\rm r} + 4H\rho_{\rm r} = 0\,, \label{eq:a_cont_rad} \\
	&& \dot{\rho}_{\rm m} + 3H\rho_{\rm m} = \alpha H \rho_{\rm d}\,, \label{eq:a_cont_dm_rad} \\
	&& \dot{\rho}_{\rm d} + 3H(1+w(a))\rho_{\rm d} = -\alpha H \rho_{\rm d}\,. \label{eq:a_cont_de_rad}
\end{eqnarray}

From the radiation continuity equation \eqref{eq:a_cont_rad}, using the conversion $d/dt = H(d/d\ln a)$, we directly obtain:
\begin{equation}
	\rho_{\rm r}(a) = \rho_{\rm r,0} \, a^{-4}\,. \label{eq:a_rho_r_sol}
\end{equation}

For dark energy, substituting $w(a)$ into Eq.~\eqref{eq:a_cont_de_rad} and converting the time derivative to a derivative with respect to $a$ ($\dot{\rho}_{\rm d} = aH d\rho_{\rm d}/da$), we have:
\begin{equation}
	\frac{d\rho_{\rm d}}{\rho_{\rm d}} = - \left[ \frac{3(1+w_0+w_a) + \alpha}{a} - 3w_a \right] da\,.
\end{equation}
Integrating this from the present day ($a=1$) to an arbitrary scale factor $a$ yields:
\begin{equation}
	\rho_{\rm d}(a) = \rho_{\rm d,0} \, a^{-[3(1+w_0+w_a) + \alpha]} \, e^{3w_a(a-1)}\,. \label{eq:a_rho_d_sol_rad}
\end{equation}

For dark matter, Eq.~\eqref{eq:a_cont_dm_rad} can be rewritten as a first-order linear differential equation:
\begin{equation}
	\frac{d\rho_{\rm m}}{da} + \frac{3}{a}\rho_{\rm m} = \frac{\alpha}{a}\rho_{\rm d}(a)\,.
\end{equation}
Multiplying both sides by the integrating factor $a^3$, we get:
\begin{equation}
	\frac{d}{da} \left( a^3 \rho_{\rm m} \right) = \alpha a^2 \rho_{\rm d}(a)\,.
\end{equation}
Substituting Eq.~\eqref{eq:a_rho_d_sol_rad} into the above equation and integrating from $1$ to $a$, we obtain:
\begin{equation}
	\rho_{\rm m}(a) = \rho_{\rm m,0} a^{-3} + \alpha \rho_{\rm d,0} a^{-3} e^{-3w_a} \int_1^a x^{-(1+3w_0+3w_a+\alpha)} e^{3w_a x} dx\,. \label{eq:a_rho_m_sol_rad}
\end{equation}
Note that for the $w$CDM model ($w_a=0, w_0=w$), the integral can be solved analytically, yielding the exact form used in the main text.

\subsection{Derivation of $E(z)^2$, $\Omega_{\rm m}(z)$, and $d\Omega_{\rm m}/d\ln a$}
The normalized Hubble parameter $E(a)^2 \equiv H(a)^2/H_0^2$ is given by:
\begin{equation}
	E(a)^2 = \frac{\rho_{\rm m}(a) + \rho_{\rm d}(a) + \rho_{\rm r}(a)}{\rho_{\rm crit,0}}\,.
\end{equation}
where $\rho_{\rm crit,0} = 3H_0^2 / (8\pi G)$ is the present-day critical density of the universe. Substituting Eqs.~\eqref{eq:a_rho_r_sol}--\eqref{eq:a_rho_m_sol_rad} into this expression, using $\Omega_{i,0} = \rho_{i,0}/\rho_{\rm crit,0}$, and applying the redshift relation $a = (1+z)^{-1}$, we obtain the full expression for $E(z)^2$:
\begin{eqnarray}
	E(z)^2 &=& \Omega_{\rm r,0}(1+z)^4 + \Omega_{\rm m,0}(1+z)^3 + \Omega_{\rm d,0}(1+z)^{3(1+w_0+w_a)+\alpha} e^{3w_a(\frac{1}{1+z}-1)} \nonumber \\
	&& + \alpha \Omega_{\rm d,0} (1+z)^3 e^{-3w_a} \int_1^{\frac{1}{1+z}} x^{-(1+3w_0+3w_a+\alpha)} e^{3w_a x} dx \,.
\end{eqnarray}

The fractional energy density of matter is defined as $\Omega_{\rm m}(a) = \rho_{\rm m}(a)/\rho_{\rm crit}(a)$. Since the time-dependent critical density is $\rho_{\rm crit}(a) = \rho_{\rm crit,0} E(a)^2$, we can express $\Omega_{\rm m}(a)$ as:
\begin{equation}
	\Omega_{\rm m}(a) = \frac{\rho_{\rm m}(a)}{\rho_{\rm crit,0} E(a)^2}\,.
\end{equation}
Substituting Eq.~\eqref{eq:a_rho_m_sol_rad} into this definition and converting to redshift $z$, we obtain the exact expression for the interacting CPL model:
\begin{equation}
	\Omega_{\rm m}(z) = \frac{\Omega_{\rm m,0}(1+z)^3 + \alpha \Omega_{\rm d,0} (1+z)^3 e^{-3w_a} \int_1^{\frac{1}{1+z}} x^{-(1+3w_0+3w_a+\alpha)} e^{3w_a x} dx}{E(z)^2}\,.
\end{equation}
For the interacting $w$CDM model ($w_a=0, w_0=w$), the integral evaluates to $[x^{-(3w+\alpha)}]_1^{1/(1+z)} / -(3w+\alpha)$, which simplifies to the analytical form presented in Eq.~\eqref{omegam_wcdm} of the main text.

To derive the differential equation for $\Omega_{\rm m}$, we take the time derivative of $\Omega_{\rm m} = \rho_{\rm m}/\rho_{\rm crit}$:
\begin{equation}
	\dot{\Omega}_{\rm m} = \frac{\dot{\rho}_{\rm m}}{\rho_{\rm crit}} - \Omega_{\rm m} \frac{\dot{\rho}_{\rm crit}}{\rho_{\rm crit}} = -3H\Omega_{\rm m} + \alpha H \Omega_{\rm d} - 2\Omega_{\rm m} \frac{\dot{H}}{H}\,.
\end{equation}
Taking the time derivative of the Friedmann equation gives:
\begin{equation}
	\frac{\dot{H}}{H^2} = -\frac{3}{2}\left(\Omega_{\rm m} + \Omega_{\rm d}(1+w(a)) + \frac{4}{3}\Omega_{\rm r}\right)\,.
\end{equation}
Substituting this into the $\dot{\Omega}_{\rm m}$ equation and using the flatness condition $\Omega_{\rm m} + \Omega_{\rm d} + \Omega_{\rm r} = 1$ to replace the $\Omega_{\rm m}$ inside the parenthesis, we have:
\begin{eqnarray}
	\dot{\Omega}_{\rm m} &=& -3H\Omega_{\rm m} + \alpha H \Omega_{\rm d} + 3H\Omega_{\rm m} \left( 1 + w(a)\Omega_{\rm d} + \frac{1}{3}\Omega_{\rm r} \right) \nonumber \\
	&=& H(\alpha \Omega_{\rm d} + 3w(a)\Omega_{\rm m}\Omega_{\rm d} + \Omega_{\rm m}\Omega_{\rm r})\,.
\end{eqnarray}
Converting the time derivative to a derivative with respect to $\ln a$, we obtain the final dynamical equation:
\begin{equation}
	\frac{d\Omega_{\rm m}}{d\ln a} = \alpha \Omega_{\rm d} + 3w(a)\Omega_{\rm m}\Omega_{\rm d} + \Omega_{\rm m}\Omega_{\rm r}\,.
\end{equation}

\section{Compilation of $f\sigma_8$ Measurements}
\label{app:fs8_data}

In this appendix, we provide the complete compilation of the observed growth rate data, $f\sigma_8(z)$, derived from various redshift-space distortion (RSD) measurements used in our analysis. The primary dataset consists of 20 measurements spanning the redshift range $0.001 \leq z \leq 1.48$, as summarized in Table~\ref{BAOfs8}. These values correspond to the standard consensus measurements (including the combined BAO+RSD fits for the SDSS data) and are used when performing the joint analysis with the SDSS BAO dataset (SHCB$_{\mathrm{SDSS}}$+RSD).

\begin{table}[htp]\center
	\begin{tabular}{c c c c c}
		\hline\hline
		\ \ \   Index\ \ \ \ \  & \ \ \ \ \ Dataset \ \ \ \ & \ \ Redshift  & \ \ \ \ \  $f\sigma_8(z)$ \ \ \ \ \ \ \ &  \ \ References \\
		\hline
		1 & 2MTF & 0.001  &  $0.505\pm0.085$  & ~\cite{Howlett2017} \\
		2 & ALFALFA & 0.013  &  $0.46\pm0.06$  & ~\cite{Avila2021} \\
		3 & 6dFGS+SnIa & 0.02  &  $0.428\pm0.0465$  & ~\cite{Huterer2017} \\
		4 & SNeIa+IRAS & 0.02  &  $0.398\pm0.065$  & ~\cite{Turnbull2012,Hudson2012} \\
		5 & 2MASS & 0.02  &  $0.314\pm0.048$  & ~\cite{Hudson2012,Davis2011} \\
		6 & 2dFGRS & 0.17  &  $0.51\pm0.06$  & ~\cite{Song2009} \\
		7 & GAMA & 0.18  &  $0.36\pm0.09$  & ~\cite{Simpson2016} \\
		8 & GAMA & 0.38  &  $0.44\pm0.06$  & ~\cite{Simpson2016} \\
		9 & WiggleZ & 0.44  &  $0.413\pm0.08$  & ~\cite{Blake2012} \\
		10 & WiggleZ & 0.60  &  $0.39\pm0.063$  & ~\cite{Blake2012} \\
		11 & WiggleZ & 0.73  &  $0.437\pm0.072$  & ~\cite{Blake2012} \\
		12 & Vipers PDR-2 & 0.60  &  $0.55\pm0.12$  & ~\cite{Pezzotta2017} \\
		13 & Vipers PDR-2 & 0.86  &  $0.40\pm0.11$  & ~\cite{Pezzotta2017} \\
		14 & FastSound & 1.40  &  $0.482\pm0.116$  & ~\cite{Okumura2016} \\
		15 & SDSS-MGS & 0.15  &  $0.530\pm0.160$  & ~\cite{Howlett2015} \\
		16 & SDSS-BOSS-Galaxy & 0.38  &  $0.497\pm0.045$  & ~\cite{AlamShadab2017} \\
		17 & SDSS-BOSS-Galaxy & 0.51  &  $0.459\pm0.038$  & ~\cite{AlamShadab2017} \\
		18 & SDSS-eBOSS-LRG & 0.70  &  $0.473\pm0.041$  & ~\cite{Bautista2021,Gil2020} \\
		19 & SDSS-eBOSS-ELG & 0.85  &  $0.315\pm0.095$  & ~\cite{De2021,Tamone2020} \\
		20 & SDSS-eBOSS-Quasar & 1.48  &  $0.462\pm0.045$  & ~\cite{Neveux2020,Hou2021} \\
		\hline
	\end{tabular}
	\caption{Compilation of the observed growth rate in the form of $f\sigma_8(z)$ derived from RSD measurements. For the SDSS data (indices 15--20), these values represent the combined BAO+RSD consensus measurements.}
	\label{BAOfs8}
\end{table}

It is important to note the specific treatment of the SDSS measurements (indices 15--20) in our MCMC analysis when the DESI DR2 BAO dataset is employed. To strictly avoid double-counting the BAO information when combining with DESI BAO (SHCB$_{\mathrm{DESI}}$+RSD), we must decouple the RSD measurements from the BAO measurements.

For these six SDSS data points, we adopt the ``RSD-only'' consensus measurements from the final SDSS-IV release\footnote{\url{https://www.sdss4.org/science/final-bao-and-rsd-measurements/}}. The exact values used for the DESI-based analysis are explicitly listed in Table~\ref{tab:rsd_only}. Consequently, the likelihood evaluation for these points relies purely on their RSD constraints, independent of the background BAO distances.

\begin{table}[htp]\center
	\begin{tabular}{c c c c}
		\hline\hline
		\ \ \   Index\ \ \ \ \  & \ \ \ \ \ Dataset \ \ \ \ & \ \ Redshift  & \ \ \ \ \  $f\sigma_8(z)$ (RSD-only) \ \ \ \ \ \ \ \\
		\hline
		15 & SDSS-MGS & 0.15  &  $0.530\pm0.160$  \\
		16 & SDSS-BOSS-Galaxy & 0.38  &  $0.500\pm0.047$  \\
		17 & SDSS-BOSS-Galaxy & 0.51  &  $0.455\pm0.039$  \\
		18 & SDSS-eBOSS-LRG & 0.70  &  $0.448\pm0.043$  \\
		19 & SDSS-eBOSS-ELG & 0.85  &  $0.315\pm0.095$  \\
		20 & SDSS-eBOSS-Quasar & 1.48  &  $0.462\pm0.045$  \\
		\hline
	\end{tabular}
	\caption{The ``RSD-only'' consensus measurements for the six SDSS data points, used specifically in the SHCB$_{\mathrm{DESI}}$+RSD joint analysis to avoid double-counting BAO information.}
	\label{tab:rsd_only}
\end{table}

\clearpage


\end{document}